\documentclass[12pt,preprint]{aastex}
\begin{document} 

\title{Where the Blue Stragglers Roam: Searching for a Link Between
  Formation and Environment}
\author{Nathan Leigh, Alison Sills}
\affil{Department of Physics and Astronomy, McMaster University, 1280
  Main Street West, Hamilton, ON, L8S 4M1, Canada}
\email{leighn@mcmaster.ca, asills@mcmaster.ca}
\author{Christian Knigge}
\affil{School of Physics and Astronomy, Southampton University,
  Highfield, Southampton, SO17 1BJ, UK}
\email{christian@astro.soton.ac.uk}

\begin{abstract}

  The formation of blue stragglers is still not completely understood,
  particularly the relationship between formation environment and
  mechanism.  We use a large, homogeneous sample of blue stragglers in
  the cores of 57 globular clusters to investigate the relationships
  between blue straggler populations and their environments.  We use a
  consistent definition of ``blue straggler'' based on position in the
  color-magnitude diagram, and normalize the population relative to
  the number of red giant branch stars in the core. We find that the
  previously determined anti-correlation between blue straggler
  frequency and total cluster mass is present in the purely core
  population.  We find some weak correlations with central velocity
  dispersion and with half-mass relaxation time. The blue straggler
  frequency does not show any trend with any other cluster
  parameter. Even though collisions may be expected to be a dominant
  blue straggler formation process in globular cluster cores, we find
  no correlation between the frequency of blue stragglers and the
  collision rate in the core.  We also investigated the blue straggler
  luminosity function shape, and found no relationship between any
  cluster parameter and the distribution of blue stragglers in the
  color-magnitude diagram. Our results are inconsistent with some
  recent models of blue straggler formation that include collisional
  formation mechanisms, and may suggest that almost all observed blue
  stragglers are formed in binary systems.

\end{abstract}


\section{Introduction} \label{intro}
Blue stragglers are stars that are brighter and bluer (hotter) than
the main-sequence turn-off.  Having seemingly missed their peers' cue
to make the transition to lower temperatures, stars having their same
mass have already evolved off the MS and begun their ascent up the
giant branch (GB).  First discovered by \cite{sandage53} in the
cluster M3, blue straggler stars (BSSs) are an example of the
inability of standard stellar evolution alone to explain all stars,
and are used as the prime example of the complex interplay between
stellar evolution and stellar dynamics \citep[e.g.][]{sills05}.
Numerous formation mechanisms have been proposed over the years, but
the currently favored mechanisms are thought to depend on cluster
dynamics.

There is a consensus that blue stragglers are the products of
stellar mergers between two (or more) low mass MS stars, either
through direct stellar collisions or the coalescence of a 
binary system \citep{leonard89, livio93, stryker93, bailyn95}.  In
order for a binary system to coalesce, Roche lobe overflow must occur,
triggering mass transfer from the outer envelope of an evolved donor
onto that of its companion.  As such, the process is dependent on the
donor's evolutionary state.  Collisions, on the other hand, do not
depend as much on the evolutionary status of the participants since in
this case two (or more) stars pass very close to one another, form a
brief and highly eccentric binary system, and then rapidly spiral
inwards and merge as tidal forces dissipate the orbital energy.

There is evidence to suggest that both formation mechanisms do occur,
though the preferred creation pathway appears dependent on the cluster
environment \citep[e.g.][]{warren05, mapelli06}. Observations from the
Hubble Space Telescope (HST) indicate that BSSs are centrally
concentrated in globular clusters \citep{ferraro99}, though they have
been detected throughout all clusters that have been surveyed
well. The BSS populations have been found to have a bimodal radial
distribution in clusters like M55 \citep{zaggia97}, M3
\citep{ferraro97}, and 47 Tuc (NGC 104) \citep{ferraro04}, with
elevated numbers in the cores followed by a "zone of avoidance" at a
few core radii and a final rise in BS numbers towards the cluster
outskirts. This bimodal trend is thought to arise because two separate
formation mechanisms are dominating in the core and in the cluster
periphery, with mass transfer predominantly taking place in the outer,
less dense regions and collisions mainly occurring towards the cluster
center.  In support of this last point, eclipsing binaries consisting
of main sequence components having short periods and sharing a common
envelope, called contact or W UMa binaries, have been observed among
BSSs in globular \citep{mateo90, yan94} and open clusters
\citep{ahumada95}.

There is an additional complication, however, when considering the
effects of stellar collisions.  It has been suggested that collisions
need not only occur between two single stars in exceptionally dense
environments, but rather might also occur in less dense systems via
resonant interactions between primordial binaries \citep{bacon96}.
Two binary pairs locked in a tightly bound 4-body system can actually
increase the rate of collisions by increasing the collisional
cross-section of the system \citep{fregeau04}.  Indeed, multiple
collisions are thought to be responsible for some of the blue
stragglers we see, in particular those having masses around twice that
of our Sun or more \citep{sepinsky00}, or the unusual blue straggler
binary system in 47 Tucanae which probably requires 3 progenitor stars
\citep{knigge06}. Clearly both cluster dynamics and binary star
populations will determine how many of these binary-mediated
collisions will occur.

Recently, \cite{piotto04} examined the CMDs of 56 different GCs,
comparing the BSS frequency to cluster properties like total mass
(absolute luminosity) and central density.  The relative frequencies
were approximated by normalizing the number of BSSs to the HB or the
red giant branch (RGB), though the results did not depend on which
specific frequency they chose.  They found that the most massive
clusters had the lowest frequency of BSSs, and that there was little
or no correlation between BSS frequency and cluster collisional
parameter.  They also showed that the BSS luminosity function for the
most luminous clusters had a brighter peak and extended to brighter
luminosities than did that of the fainter clusters.

The absence of a correlation between BSS frequency and collision
number, in particular, is surprising, since other evidence suggests
that dynamical interactions do affect stellar populations in GCs. For
example, GCs host enhanced numbers of unusual short-period binary
systems such as low-mass X-ray binaries (LMXBs; \citep{dieball05}),
cataclysmic variables (CVs; \citep{knigge03}), and millisecond
pulsars (MSPs; \citep{edmonds03}). These objects, like blue
stragglers, are thought to trace the dynamically-created populations
of clusters. Their presence has been linked to the high densities
found in the cores of GCs, which are thought to lead to an increase in
the frequency of close encounters and thus in the formation rate of
exotic binary systems.  Indeed, the number of close binaries in GCs
observed in X-rays has been shown to be correlated with the predicted
stellar encounter rate of the cluster \citep{pooley03}.

A useful quantity for parameterizing the surface brightness
distribution of GCs is the core radius, $r_c$, defined as the distance from
the cluster center at which the surface brightness is half its
central value. That is, at a distance of $r_c$ from
the center of a King model globular cluster, the density is expected
to have dropped off to around a third of the density at the cluster
center \citep{spitzer87}. This then implies that the core encloses the
densest regions of the cluster by at least a factor of a few and hence
one would expect interactions between stars, specifically collisional
processes, to occur with the greatest frequency therein.

In our quest to determine the cluster environment's effect on BSS
formation, we decided to focus solely on the stars found within the
core in order to isolate the ones most likely to undergo stellar
encounters. We wanted to search for empirical evidence for collisional BSS
formation, and the cores of clusters are the most likely place for
collisions to not only occur, but also to dominate \citep{mapelli06,
ferraro04}. Previous attempts to connect blue straggler populations to
global cluster properties \citep{piotto04,davies04} did not attempt to
focus on a single, homogeneous environment. By concentrating solely on
the core, we therefore maintain consistent sampling from cluster to
cluster. We note that while directing our attention to the core allows
us to isolate an approximately uniform dynamical environment, it also
presents a statistical complication in post-core collapse clusters
since these tend to have small core radii and the star counts therein
thus tend to be restricted.  Fortunately, only a small fraction of the
clusters used in this paper have undergone core collapse and so this
effect should not have a significant impact on our results.

In this paper, we use Hubble Space Telescope (HST) data to look for
possible trends between relative BSS frequency and various cluster
parameters, including total cluster mass, central density, central
velocity dispersion as well as collisional parameters.  We discuss our
dataset, as well as our methodology for BSS selection in the cluster
CMDs in Section~\ref{database}.  In Section~\ref{results} we present
our results, including trends in the relative BSS frequencies as well
as a comparison of blue straggler luminosity functions (BSLFs).  We
summarize and discuss our findings in Section~\ref{discussion}.

\section{The Data} \label{database}

The color-magnitude diagrams and photometric databases for 74
Galactic globular clusters were used in this paper.  The observations,
taken from \cite{piotto02}, were made using the HST's WFPC2 camera in
the F439W and F555W bands, with the PC camera centered on the cluster
center in each case.  Generally, the field of view for each cluster
contains anywhere from a few thousand to roughly 47,000 stars.  Of the
74 potential GGCs, only 57 were deemed fit for analysis, with the
remaining 17 having been discarded due to poor reliability of the data
at or above the main sequence turn-off based on the overall appearance
of their respective CMDs. The positions of the stars, as
well as their magnitudes in both the F439W and F555W bands and the B
and V standard Johnson system, can be found at the Padova Globular
Cluster Group Web pages at http://dipastro.pd.astro.it/globulars.
Core radii and other cluster parameters were taken from
\citep{harris96} and \citep{pryor93}.

The data available at the Padova website have not been corrected for
completeness. This could be a problem for our analysis if we cannot
accurately determine the total number of blue stragglers, red giant,
horizontal branch, and extended horizontal branch stars. However, we
do not expect this to be an issue in this paper. All the clusters that
we retained in our sample have clearly-defined main sequence turn-offs
and sub-giant branches. The populations of interest are brighter than
the turn-off, by definition, and should be less affected by photometric
errors and completeness. We expect that the corrections for the
faintest objects that we identify as blue stragglers should be on the
order of one star or less, which will not change the results of this
paper. To be safe, however, stars having sufficiently large errors in
$m_{555}$ and $m_{439}$, respectively denoted by $\sigma_{f555W}$ and
$\sigma_{f555W}$, were also rejected from our counts if their total
error was more than 0.1 magnitudes.

We defined a set of boundaries in the color-magnitude diagram for
each of our populations of interest: blue stragglers, red giant branch
stars, horizontal branch stars, and extended horizontal branch
stars. The details of these definitions can be found in the appendix.
The boundaries of our blue straggler selection box were ultimately
chosen for consistency. By eliminating potential selection effects
such as ``by eye'' estimates, we were able to minimize the possibility
of counting EHB or MSTO stars as BSSs. Moreover, since we are
considering relative frequencies and there is a possibility of
mistakenly including stars other than BSSs in our counts, it seems
prudent that we at least make the attempt to systematically chose
stars in all clusters. We are most interested in using the evolved
populations to normalize the number of blue stragglers, both to give a
sense of photometric error and to remove the obvious relation that
clusters with more stars have more blue stragglers.  Therefore, we
limited ourselves to red giants with the same luminosities as the blue
stragglers. These boundaries are shown for NGC 5904 in
Figure~\ref{fig:NGC5904_CMD}. The full data for all clusters is given
in the appendix.
\begin{figure}[tbp]
\centering
\includegraphics[width=\columnwidth]{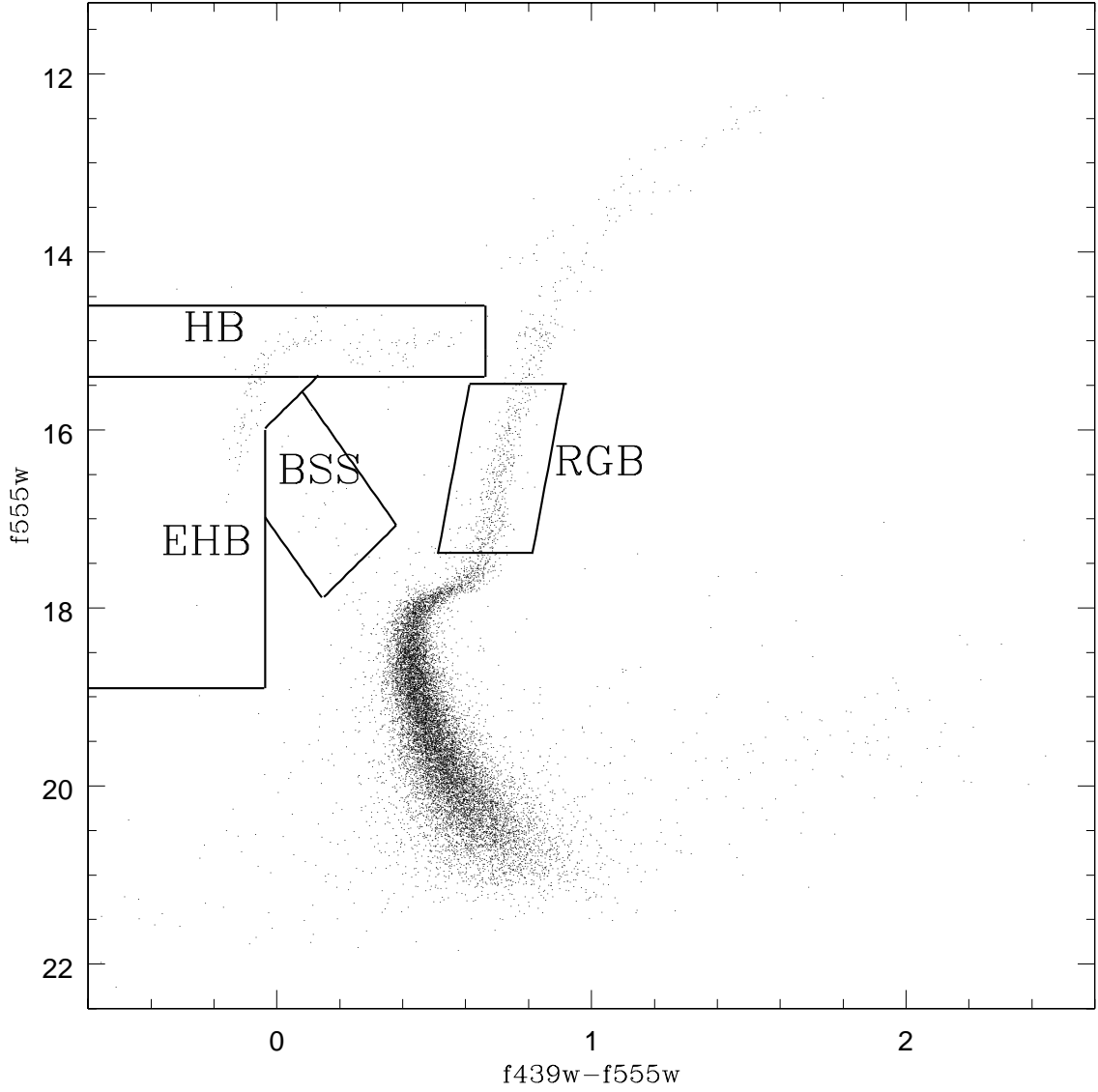}
\caption[CMD of NGC 5986]{CMD for the cluster NGC 5904 with RGB, BS, HB,
  and EHB boundaries overlaid.  
\label{fig:NGC5904_CMD}}
\end{figure}
\section{Methods of Normalization} \label{normalization}

With our sample of core BS, HB, EHB and RGB stars established consistently
from cluster to cluster, it was then necessary to address the issue of
normalization.  As one might expect, clusters having more stars tend
to be home to a larger population of BSSs. Number counts of stars
therefore had to be converted into relative frequencies.  Previously,
this was done by dividing the number of BSSs by either the number of
horizontal branch stars \citep{piotto04}:

\begin{equation}
\label{eqn:bssfreqhb}
F^{HB}_{BSS} = \frac{N_{BSS}}{N_{HB}},
\end{equation}

or by the total cluster mass \citep{demarchi06}:

\begin{equation}
\label{eqn:bssfreqmass}
F^{M_{tot}}_{BSS} = \frac{N_{BSS}}{M_{tot}}
\end{equation} 

\cite{piotto04} also looked at using the number of RGB stars for
normalization but after finding similar results, decided to simply use
the HB. The preferred means of normalization is a matter of some debate and
hence we similarly calculated relative frequencies in a few separate ways in
order to gauge which is best. Frequencies were normalized using the
HB, the EHB, the HB \& the EHB, and the RGB:

\begin{eqnarray}
\label{eqn:bssfreqall}
F^{HB}_{BSS} &=& \frac{N_{BSS}}{N_{HB}} \\
F^{EHB}_{BSS} &=& \frac{N_{BSS}}{N_{EHB}} \\
F^{HB+EHB}_{BSS} &=& \frac{N_{BSS}}{N_{HB}+N_{EHB}} \\
F^{RGB}_{BSS} &=& \frac{N_{BSS}}{N_{RGB}}
\end{eqnarray}  

Plotting these frequencies against the total V magnitude, previously
shown to exhibit a clear anti-correlation \citep{piotto04}, proved
that using the RGB gives us the tightest relationship. This reduction
of scatter was similarly observed upon comparing the BSS frequency to
other cluster parameters like the central surface brightness, the
central velocity dispersion, as well as the collisional rate.
Figure~\ref{fig:BSSnorm_round2} shows the BSS frequency versus the
total absolute V magnitude of the cluster for all four normalization
methods.
\begin{figure}[tbp]
\centering
\includegraphics[width=\columnwidth]{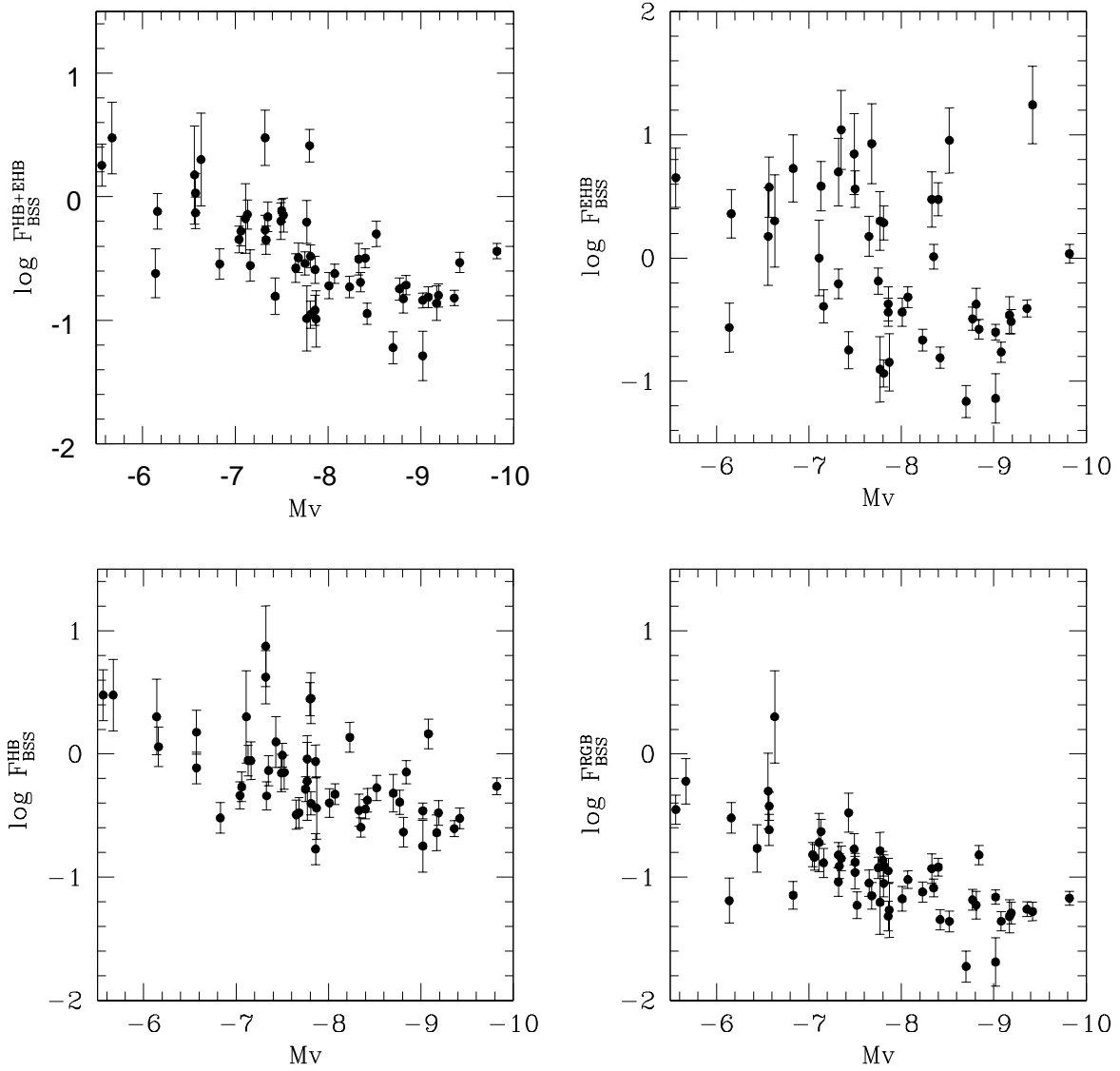}
\caption[Mv versus BSS Frequency]{Plots of the core BSS frequency
  versus the total cluster V magnitude.  Frequencies were normalized
  using RGB stars (bottom right), HB stars (bottom left), EHB stars
  (top right), and finally, HB \& EHB stars combined (top left). 
\label{fig:BSSnorm_round2}}
\end{figure}
\section{Results }\label{results}

Having obtained the numbers and frequencies of blue stragglers in the
cluster cores, we attempted to determine correlations between the blue
straggler frequency and global cluster properties such as total mass,
velocity dispersion, core density, etc. If the preferred BSS formation
mechanism in the core is direct stellar collisions, then we should see
a link between clusters with higher instances of collisions and more
pronounced BSS populations; if the formation mechanism is not
collisions, we would still expect to see a relationship between the
properties of the cluster and its stellar populations.

The central surface brightness and the core radius were taken from
\cite{harris96}.  Values for the central velocity dispersion were
taken from \cite{pryor93}.  Any other parameters, including the
central density and the total absolute luminosity, came from
\cite{piotto02}, with the exception of the normalized cluster ages
which were taken from \cite{deangeli05} and for which the Zinn \& West
(1984) metallicity scale values were employed.  Error bars for all
plots were calculated assuming Poisson statistics.

It should be noted here that while the number of blue straggler stars
in a core is a tracer of the total number of stars in the core, as
well as of the total number of HB and RGB stars, we found no
correlation between the number of blue stragglers and the number of
EHB stars. It has been speculated \citep{ferraro03} that there might
be a connection between BSS and EHB populations. The trend in that
paper (of 6 clusters) was that clusters either had bright blue
stragglers or EHB stars, but not both. With this larger
self-consistent sample, we do not find the same result. The number of
EHB stars seems to be completely independent of how many bright blue
stragglers exist in the cluster.

Relative frequencies appeared independent of the majority of the
cluster parameters analyzed, with a couple of noteworthy exceptions.
One trend observed was that the least massive clusters (those having
the lowest absolute luminosities) had the highest relative frequencies
of blue stragglers, and vice versa.  This anti-correlation was
previously observed by \cite{piotto04}, though their choice of BSS
selection and chosen method of normalization arguably led to a greater
degree of scatter in their plots. According to our analysis, using the
HB for normalization yielded correlations that were overall not as
tight as those made using the RGB. They also did not distinguish
between blue stragglers inside or outside of one core radius but
simply counted the stars in their observed fields. Given the
bimodality of the observed BSS radial distribution in some GCs, this
could have resulted in the inclusion of BSSs that were never subject to the
same dynamical conditions as those BSSs found in the core.

A similar anti-correlation was found between F$_{BSS}$ and the central
velocity dispersion, shown in
Figure~\ref{fig:BSSresults_round2_rgb}. This is perhaps unsurprising,
since velocity dispersion is known to be correlated with cluster mass
\citep{djorgovski94}. The blue straggler frequencies showed no clear
dependence on any other cluster parameters, including the central
density, the central surface brightness, and the cluster age.

Figure~\ref{fig:BSS_th_tc_rgb_round2} shows the blue straggler
frequency versus the core and half-mass relaxation times. While weak
anti-correlations were found with both the core and half-mass
relaxation times, F$_{BSS}$ was found to show a stronger
anti-correlation with the latter.  Moreover, it seems as though the
distribution begins to flatten out at higher $log t_h$, specifically
beyond around $10^9$ years. 
\begin{figure}[tbp]
\centering
\includegraphics[width=\columnwidth]{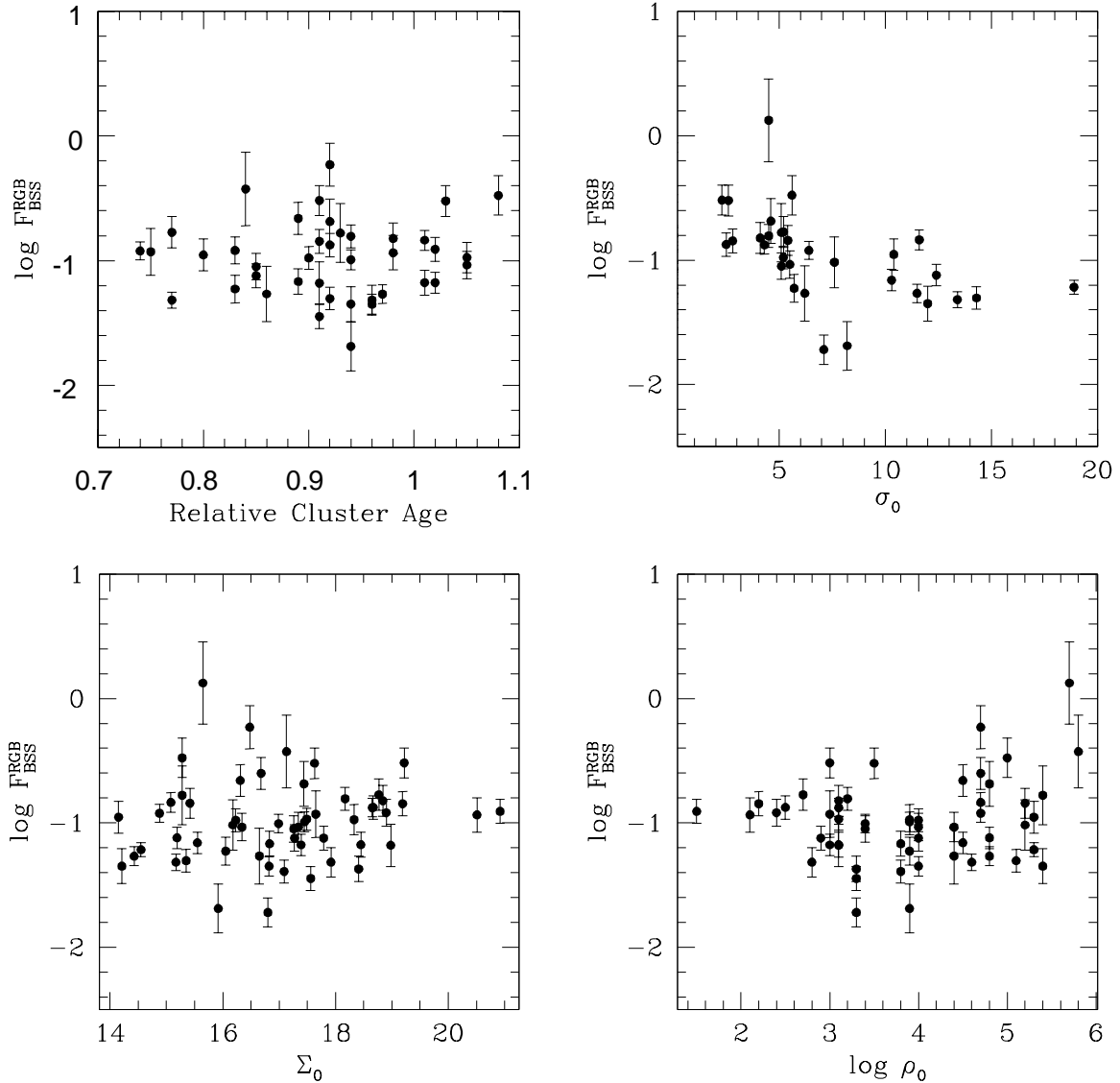}
\caption[Parameters versus BSS Frequency]{Plots of the core BSS frequency
  versus the logarithm of the central density (bottom right), the
  central surface brightness (bottom left), the relative cluster age
  (top left), and the central velocity dispersion (top right).
  Frequencies were normalized using RGB stars.  The central density is
  given in units of L$_{\odot}$ pc$^{-3}$, the central surface
  brightness in units of V mag arcsecond$^{-2}$, and the central
  velocity dispersion in units of km s$^{-1}$.  The cluster age is
  normalized, however, and its values represent the ratio between
  the cluster age and the mean age of a group of metal-poor clusters
  as described in \cite{deangeli05}.  
\label{fig:BSSresults_round2_rgb}}
\end{figure}

\begin{figure}[tbp]
\centering
\includegraphics[width=\columnwidth]{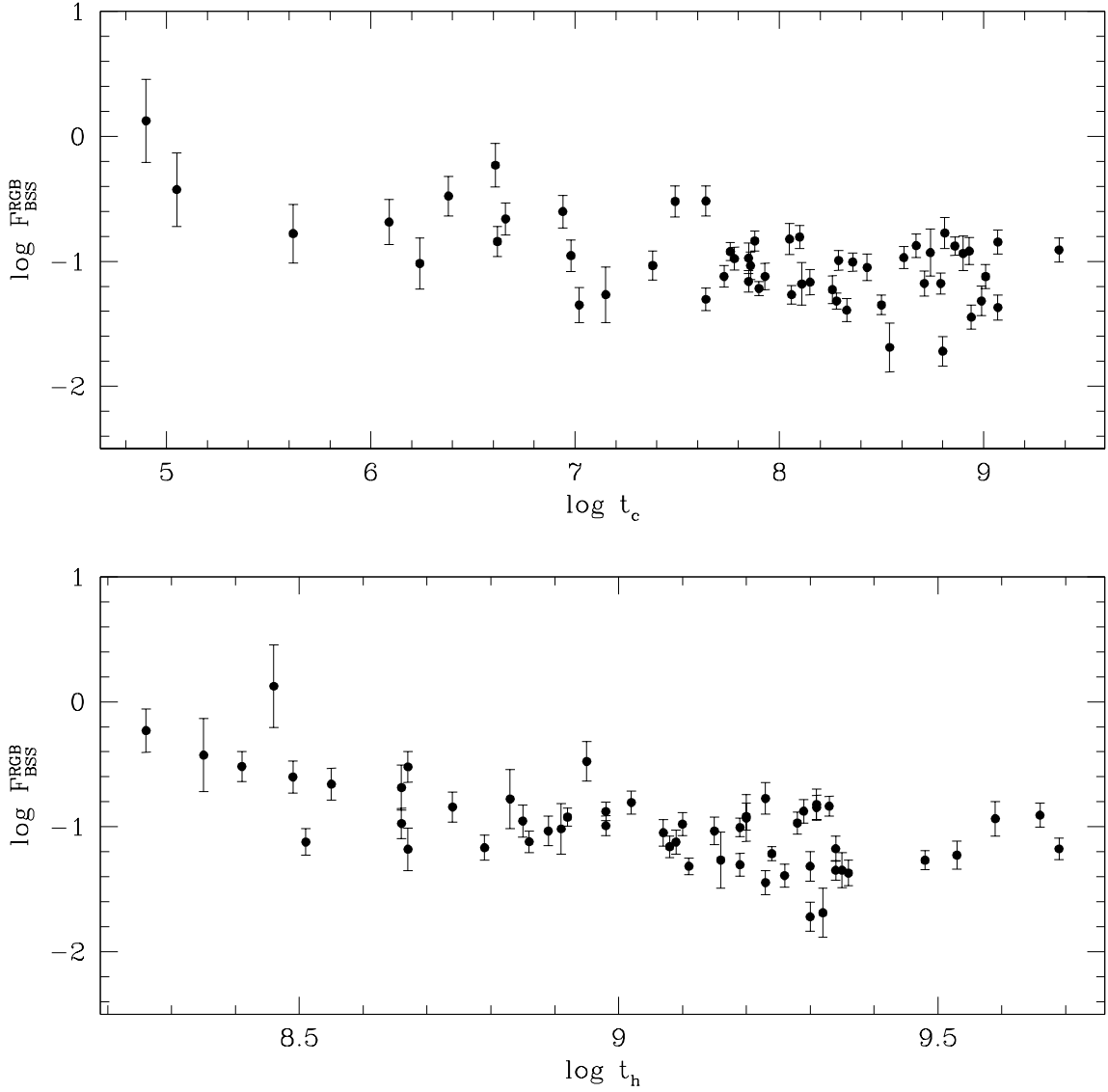}
\caption[Log of Core Relaxation Time and at the Half Mass-Radius versus BSS Frequency]{Plots of the core BSS frequency
  versus the logarithm of the core relaxation time in years (top), and
  the logarithm of the relaxation time at the half-mass radius in
  years (bottom).  Frequencies were normalized using RGB stars.  Note
  the anti-correlation that exists between F$_{BSS}$ and log $t_h$.
\label{fig:BSS_th_tc_rgb_round2}}
\end{figure}
We also considered the brightest blue stragglers (BBSSs) separately,
under the assumption that these stars are most likely to be collision
products.  Figure 2 of \cite{monkman06} shows that, in the case of 47
Tuc (NGC 104), the number of bright blue stragglers falls off
noticeably outside the cluster core.  These brightest blue stragglers
found only within the core have a B magnitude of less than about 15.60
mag, or a V magnitude of less than about 15.36 mag.  Assuming the
BBSSs in other clusters are similar to the ones found in 47 Tuc, we
defined the brightest BSSs as those having a V magnitude of 1.74 mag
brighter than that of the MSTO.  As illustrated in
Figure~\ref{fig:BBSS4x4_round2}, the usual trends with $M_V$ and the
central velocity dispersion were found.  No new trends between the
BBSS relative frequencies and any cluster parameters emerged.
\begin{figure}[tbp]
\centering
\includegraphics[width=\columnwidth]{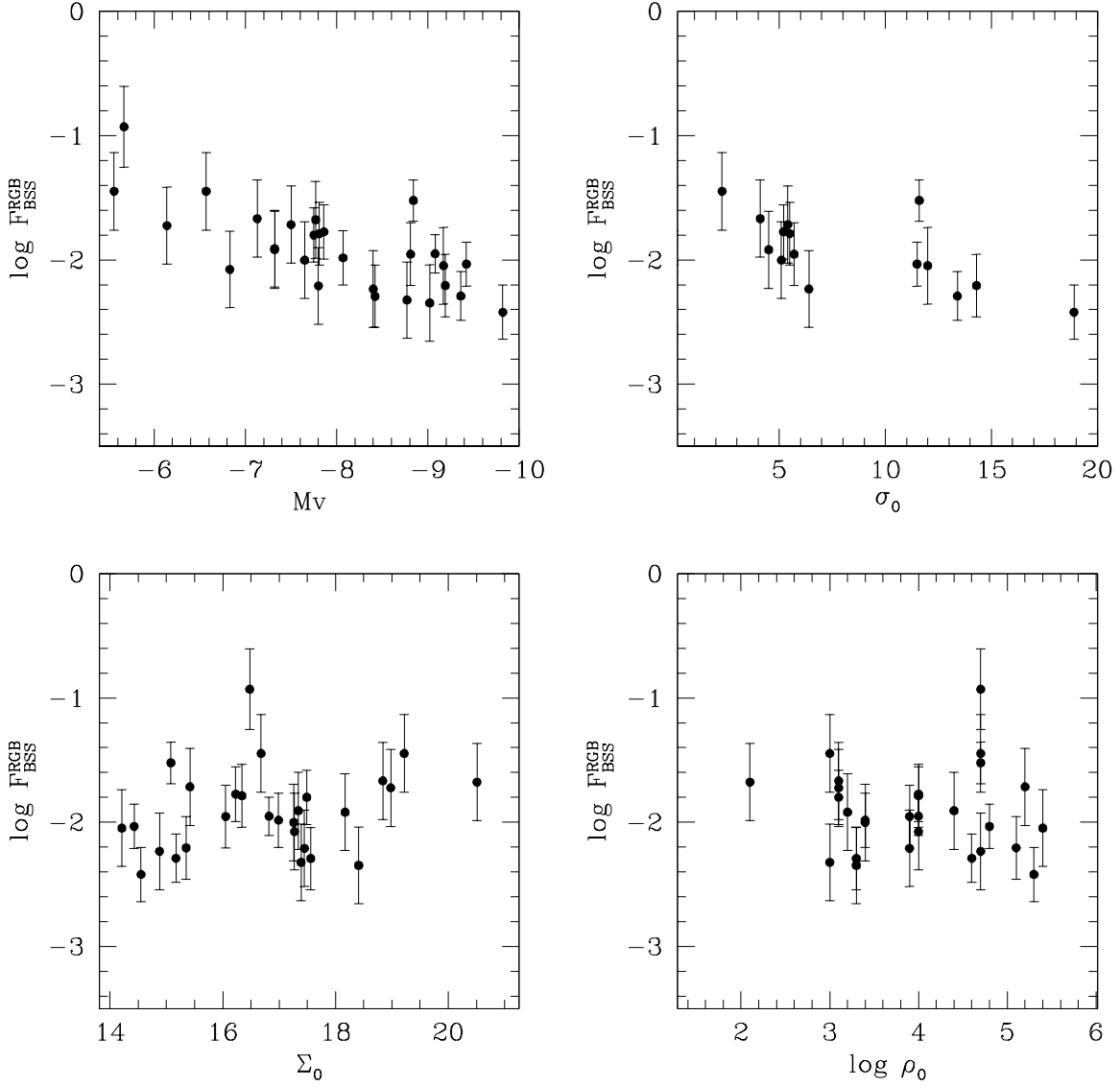}
\caption[Parameters Versus Brightest BSSs]{Plots of the brightest core
  BSS frequency versus the logarithm of the central density (bottom
  right), the central surface brightness (bottom left), the total
  cluster V magnitude (top left), and the central velocity dispersion
  (top right).  Frequencies were normalized using RGB stars.  The
  central density is given in units of L$_{\odot}$ pc$^{-3}$, the
  central surface brightness in units of V mag arcsecond$^{-2}$, the
  cluster magnitude in V mag, and the central velocity dispersion in
  units of km s$^{-1}$.
\label{fig:BBSS4x4_round2}}
\end{figure}
We also looked at the relative BSS frequencies in only the most
massive clusters for which $M_V <$ -8.8 under the assumption of
\cite{davies04} that the BSSs in these clusters should predominantly
be collision products.  Collisional BSSs are thought to be brighter
than those formed from primordial binaries.  As illustrated in
Figure~\ref{fig:BSS4x4_brightestMv_round2}, no trends were found
between BSS relative frequencies and any cluster parameters when
clusters with $M_V >$ -8.8 were ignored.  Indeed, the
previously established trend between $M_v$ and F$_{BSS}$ is
considerably weakened by eliminating clusters with $M_V >$ -8.8. 
\begin{figure}[tbp]
\centering
\includegraphics[width=\columnwidth]{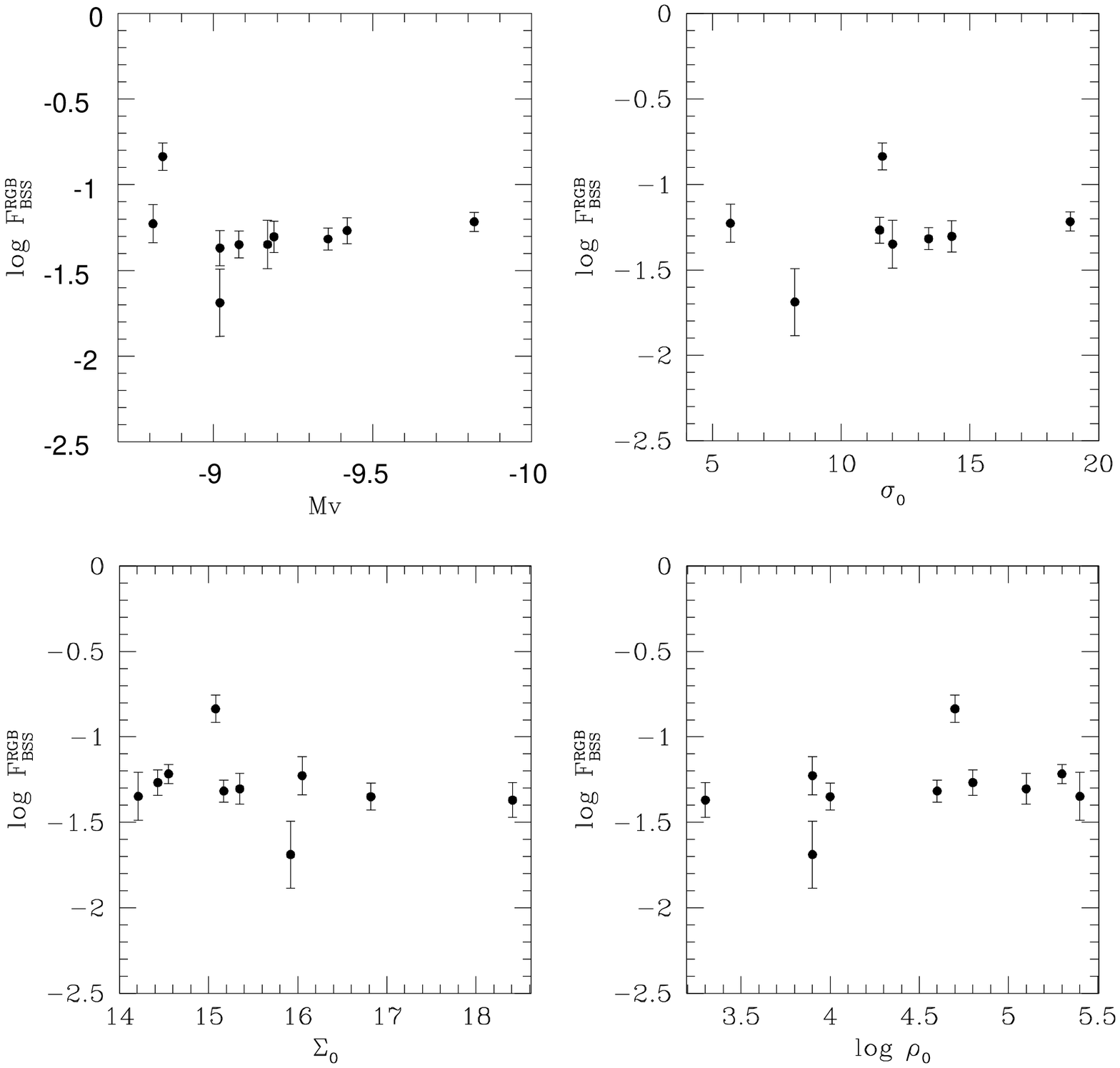}
\caption[Parameters Versus Brightest BSSs]{Plots of the core BSS
  frequency in only the brightest clusters ($M_V < -8.8$)
  versus the logarithm of the central density (bottom right), the
  central surface brightness (bottom left), the total cluster V
  magnitude (top left), and the central velocity dispersion (top
  right).  Frequencies were normalized using RGB stars.
  The central density is given in units of L$_{\odot}$ pc$^{-3}$, the
  central surface brightness in units of V mag arcsecond$^{-2}$, the
  cluster magnitude in V mag, and the central velocity dispersion in
  units of km s$^{-1}$. 
\label{fig:BSS4x4_brightestMv_round2}}
\end{figure}
BSS frequencies were also plotted against a parameter used to
approximate the rate of stellar collisions per year. Following
\citet{pooley06},

\begin{equation}
\label{eqn:collparameter2}
{\Gamma} = \frac{{\rho}_0^{2}r_c^{3}}{{\sigma}_0},
\end{equation}

where $\rho_0$ is the central density in units of L$_{\odot}$
pc$^{-3}$, $\sigma_0$ is the central velocity dispersion in km
s$^{-1}$ and $r_c$ is the core radius in parsecs. If there is a tight
correlation between the fraction of blue stragglers and $\Gamma$, then
we can conclude that direct stellar collisions are responsible for
most of the blue stragglers in cluster cores.
Figure~\ref{fig:BSS_coll_rgb_round2} shows, if anything, a decline in
BSS frequency with increasing collisional rate.  This weak
anti-correlation is likely not an artifact of the more populous
clusters having more stars available to undergo collisions since we
are dealing with normalized BSS frequencies as opposed to pure number
counts. This anti-correlation has been seen by \citet{piotto04} and
\citet{sandquist05} for blue straggler populations from a larger
region of the cluster. The trend is weak, and one could argue that
there is no correlation.
\begin{figure}[tbp]
\centering
\includegraphics[width=\columnwidth]{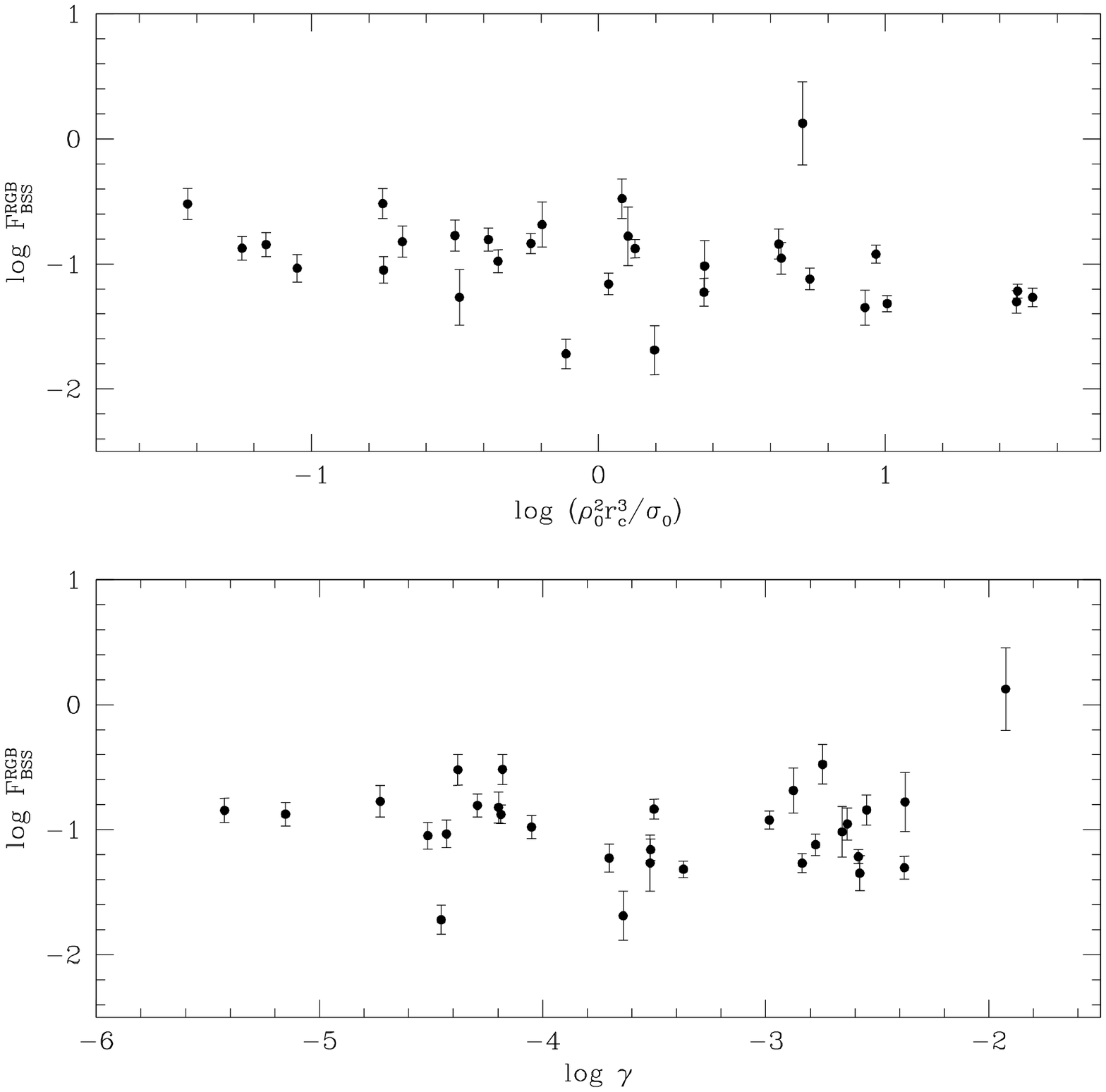}
\caption[Collisional Parameter versus BSS Frequency]{Plot of the
  BSS frequency within the cluster core versus the rate of stellar
  collisions per year using Equation~\ref{eqn:collparameter2} as the
  collisional parameter (top).  $\rho_0$ is the central density in units
  of L$_{\odot}$ pc$^{-3}$, $\sigma_0$ is the central velocity
  dispersion in km s$^{-1}$, and $r_c$ is the core radius in
  parsecs.  BSS frequency is also plotted against the probability of a
  stellar collision occurring within the core in one year (bottom).
  Frequencies were normalized using core RGB stars.
\label{fig:BSS_coll_rgb_round2}}
\end{figure}
An additional comparison can be made to the probability that a given
star will undergo a collision in one year, denoted by $\gamma$. We
divide the rate of stellar collisions by the total number of stars in
the cluster core, $N_{\rm star}$, found by directly counting them in
Piotto et al.'s (2002) database and then multiplying by the
appropriate geometrical correction factor.  In order for our counts to
be representative of the entire core, it was necessary to extrapolate
our results in the case of clusters for which only a fraction of the
core was sampled.  Therefore, the total number of stars was multiplied
by the ratio of the entire core area to that of the sampled region in
each cluster.  Figure~\ref{fig:BSS_coll_rgb_round2} clearly shows that
there is no dependence of F$_{BSS}$ on $\gamma$.  Similar results were
also found when we used the collisional parameter given in
\citet{piotto04}. We also looked for connections between both $\Gamma$
and $\gamma$ and the brightest blue stragglers, and the blue
stragglers in the brightest clusters. We found no trend in either case.

To quantify these dependences or lack thereof, we calculated the
Spearman correlation coefficients \citep{NumRec,Stats} between the
blue straggler frequency and a variety of cluster parameters. The
results are given in Table \ref {table:spearman}. The correlation
coefficient $r_s$ ranges from 0 (no correlation) to 1 (completely
correlated), or to -1 (completely anti-correlated). The third column,
labeled `Probability', gives the chance that these data are
uncorrelated. Clearly the most important anti-correlations are with
the total cluster magnitude and the central velocity dispersion (the
Spearman coefficient for $M_V$ is positive for an anti-correlation
because the magnitude scale is backwards). The anti-correlation with
half-mass relaxation time also shows up here, and the frequency of
blue stragglers in clusters may also be anti-correlated with the core
relaxation time and collision rate, although this is not conclusive
from these data.
\begin{deluxetable}{lcc}
\tablecaption{Spearman Correlation Coefficients \label{table:spearman}}
\tablehead{
\colhead{Parameter} & \colhead{$r_s$} & \colhead{Probability}\\
}
\startdata
Total cluster V magnitude  	  & 0.76 & 	 7.2E-12 \\
Central velocity dispersion  	  &-0.70 &	 1.0E-05\\
Half-mass relaxation time         &-0.53 &	 2.5E-05\\
Core relaxation time              &-0.43 & 	 1.1E-03\\
Collision rate                    &-0.41 &	 0.02 \\
Surface brightness   	          & 0.17 &	 0.20\\
Central density                   & 0.08 &	 0.58  \\
Collision probability             &-0.09 &	 0.63 \\
Age  	                          & 0.02 &       0.91  	\\
\enddata
\end{deluxetable}
\section{Blue Straggler Luminosity Functions} \label{LFs}

Having investigated the relationship between the frequency of blue
stragglers and their host cluster properties, we now turn to looking
at the details of the blue straggler population itself.  Cumulative
blue straggler luminosity functions (BSLFs) were made for all 57
clusters in our sample. The magnitude of the MSTO differs from cluster
to cluster and so, in order to correct for these discrepancies, it was
subtracted from the BSS magnitudes in each cluster.

We wanted to quantify the shape of these luminosity functions in
order to determine if there were any connections between BSLF shape
and global cluster parameters. We found that all the BSLFs could be
well-fit using a quadratic (but not a linear) function of
magnitude. Some examples of the fits are given in the top panel of
Figure \ref{fig:BSS_LF}. The fits for all clusters are
given in the bottom panel.

We looked at each of the quadratic coefficients as a function of all
of the cluster parameters, and found that the coefficients had no
dependency on any parameter, including total cluster magnitude or
metallicity. We were particularly interested in metallicity since the
BSLF is a measure of the properties of BSS stellar evolution, which
could depend on metallicity. According to these data, it does not.
\begin{figure}[tbp]
\centering
\includegraphics[width=\columnwidth]{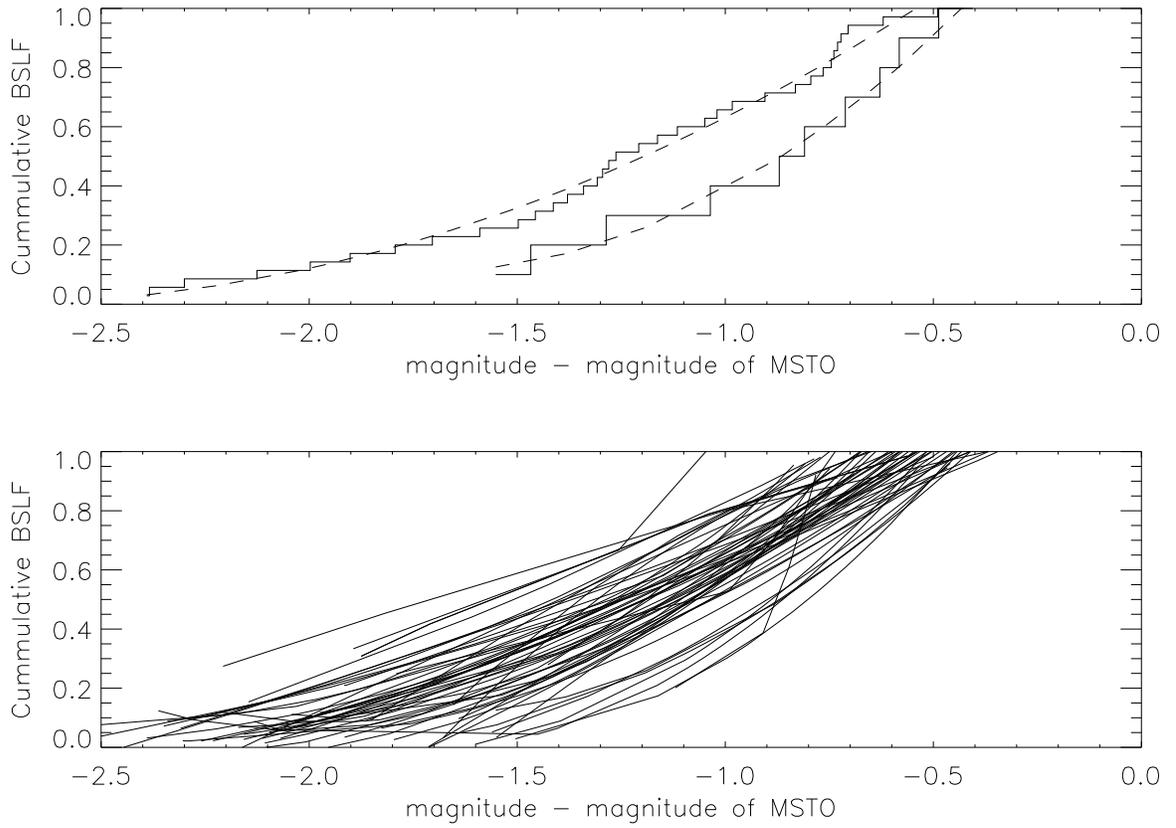}
\caption[BSLFs and fits]{Cumulative BS luminosity functions. The top
  panel shows the BSLFs for NGC 104 (47 Tucanae) and NGC 7099 (M30)
  (solid lines) along with quadratic fits to those functions (dashed
  lines). The bottom panel shows the quadratic fits to all cluster
  BSLFs. 
  \label{fig:BSS_LF}}
\end{figure}  
\citet{piotto04} argued that if the BSS formation mechanisms depend on
the cluster mass, then one would expect the blue straggler LFs to
likewise depend on the mass.  They predicted that the luminosity
distribution of collisionally produced BSSs should differ from those
created via mass transfer or the merger of a binary system due to
different resulting interior chemical profiles.  They were able to
generate separate BSLFs for clusters with $M_V <$ -8.8 and $M_V >$ -8.8
in support of their hypothesis.  Upon subtracting MSTO magnitudes from
peak BSLF magnitudes and creating individual BSLFs for clusters above and
below a total V magnitude of -8.8, we found the difference between the
two sub-sets of BSLFs to be negligible. Interestingly, there were in
total only 11 clusters in our dataset for which $M_V <$ -8.8 and so,
had we found any potential trends, their reliability would be suspect.
Any generalizations made regarding the most massive clusters should be
disregarded due to the small number of clusters in the Piotto et al.
(2002) dataset in this regime.

We repeated this experiment by binning our BSLFs according to cluster
magnitude, in bins of size 1 magnitude. The results are shown in
Figure \ref{fig:BSLF_MV_bin}. We see no trend in the peak of the BSLF
with cluster magnitude. We also tried binning the BSLFs by central
density (in bins of size 1 in $\log (\rho)$) and by half-mass
relaxation time (in bins of size 0.5 in $\log(t_h)$). Again, we found
no trend in the peak magnitude or shape of the luminosity function for any
of these parameters. We expect, given our analysis of the shapes of
the cumulative BSLFs, that binning by any other cluster parameter
will similarly yield no trends. Just to check, we performed a
Kolmogorov-Smirnov test on the luminosity functions that were binned
by absolute cluster magnitude (those shown in Figure
\ref{fig:BSLF_MV_bin}). No pairs of distributions were drawn from the
same parent distribution with more than a few percent probability. The
closest pair ($-7 < M_V < -6$ and $-9 < M_V < -8$) have a 57\%
probability of being drawn from the same distribution; and all other
pairs were below the 10\% level.
\begin{figure}[tbp]
\centering
\includegraphics[width=\columnwidth]{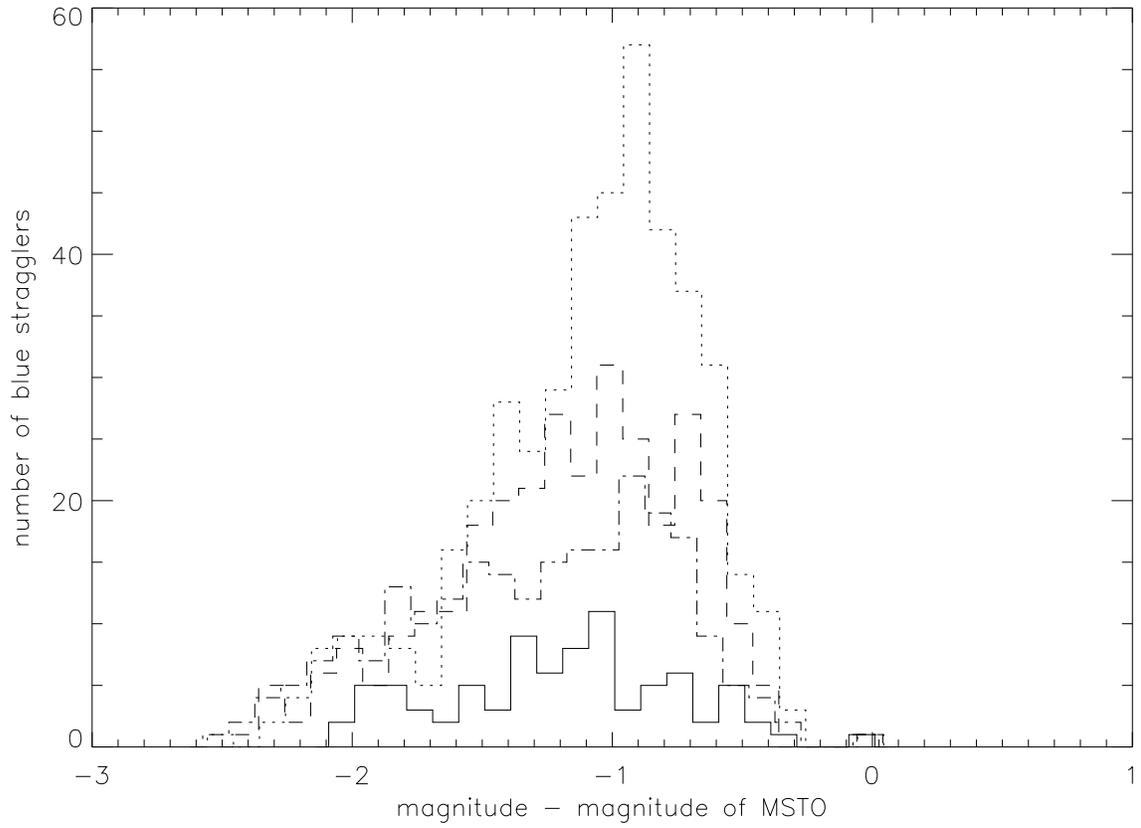}
\caption[binned BSLF with MV]{Blue straggler luminosity functions,
  binned according to total cluster magnitude. The solid line is for clusters
  with $M_V$ between -6 and -7; dotted line for clusters with $M_V$
  between -7 and -8; dashed line for clusters with $M_V$ between -8
  and -9, and the dash-dotted line is for clusters with $M_V$ between
  -9 and -10. 
\label{fig:BSLF_MV_bin}}
\end{figure}  

\section{Summary \& Discussion} \label{discussion}

We used the large homogeneous database of HST globular cluster
photometry from \citet{piotto04} to investigate correlations of blue
stragglers with their host cluster properties. First, we applied a
consistent definition of ``blue straggler'' to all our clusters. We
chose the MSTO as our starting point and defined our boundaries based
only on its location in the CMD. We also defined the location of
horizontal branch and extended horizontal branch stars in the CMD. We
looked at a variety of normalizations for our BSS frequencies before
determining that using the RGB yielded the plots with the least
scatter. 

There are disappointingly few strong correlations between the
frequency of blue stragglers in the cores of these clusters and any
global cluster parameter. We confirm the anti-correlation between the
the total integrated cluster luminosity and relative BSS frequency
found by \cite{piotto04}; we suggest an anti-correlation with central
velocity dispersion based on a high correlation coefficient comparable
to that found for the cluster luminosity. At a lower significance are
possible anti-correlations with half-mass and core relaxation times. 

The relaxation time for a cluster is a measure of how long a system
can live before individual stellar encounters become important and the
approximation of objects moving in a smooth mean potential
breaks down. Globular clusters are the quintessential collisional
dynamical systems, since their ages are typically much longer than
their relaxation times. It seems sensible that the fraction of blue
stragglers, a dynamically created population, should depend on the
relaxation times of clusters. The puzzle comes, as usual, in the
details. First, we are looking at core blue stragglers, so we might
expect the blue straggler fraction to go up with decreasing core
relaxation time. The observations do give us this kind of trend, but
not a very strong one. We might also expect the blue straggler
fraction to depend on the half-mass relaxation time, which is a better
measure of the dynamical state of the whole system. The blue straggler
fraction does depend on $t_{h}$ in the way that we expect, but only
apparently for systems with short dynamical times. Is blue straggler
formation a dynamical process which takes a while to turn on? It seems
that might be the case.

If the observed anti-correlation between F$_{BSS}$ and the central
velocity dispersion is real, then it follows that random relative
motions somehow impede stellar mergers.  A similar conclusion can be
drawn from the plots of F$_{BSS}$ versus $\Gamma$ in the event that
the speculated anti-correlations are truly representative of the
cluster dynamics.  Collisional processes therefore seem to somehow
interfere with the production of BSSs.  If the majority of BSSs are,
in fact, the remnants of coalescing binaries then it stands to reason
that an increase in the number of close encounters or collisions with
other stars could result in the disruption of a larger number of
binary systems.  On the other hand, if the majority of BSSs are the
products of stellar collisions, then it is conceivable that those
clusters having the highest collisional frequencies are the most
likely to undergo three- or four-body encounters.  As such, clusters
having the highest collisional rates could also have, on average, the
most massive BSSs resulting from an increased incidence of multi-body
mergers.  This might then contribute to the observed weak
anti-correlation between F$_{BSS}$ and the collisional rate since
clusters having a higher incidence of collisions should consequently
have a higher incidence of multi-body mergers, resulting in a
potentially lower relative BSS frequency.  We need more, and more
accurate, individual surface gravity measurements of blue stragglers
in order to explore the idea that a surplus of more massive BSSs can
be found in those clusters with a higher $\Gamma$.

F$_{BSS}$ was found to be nearly uniform with every other cluster
parameter, suggesting that all globular clusters of all properties
produce the same fraction of blue stragglers. The lack of a dependence
of F$_{BSS}$ on cluster age implies that whatever the preferred
mechanism(s) of BSS formation, it occurs in globular clusters of every
age with comparable frequencies. It should be noted that this result
does not extend to open clusters. There is a clear correlation of blue
straggler frequency with cluster age for open clusters between the
ages of 10$^8$ and 10$^{10}$ years \citep{demarchi06}.

Cumulative BSS luminosity functions were analyzed for all 57 GCs.
Unlike \cite{piotto04}, we found no real difference between the BSLFs
of the most massive clusters and those of the least massive clusters.
In fact, we found no correlation between the shape of the cumulative
luminosity function and any other cluster property. These results do
not support the notion that differing interior chemical profiles cause
collisionally-produced BSSs to have differing luminosities from those
created via mass transfer or the coalescence of primordial binaries.
It does, however, suggest that either the products of both formation
mechanisms cannot be distinguished by their luminosity functions
alone, or a single formation mechanism is operating predominantly in
all environments.

Trends were also looked for in the brightest blue stragglers, since we
suspected their enhanced brightnesses to imply a collisional
origin. We also looked at the entire core blue straggler population in
the most massive clusters (also thought to be predominantly a
collisional population).  No trends were observed. Therefore, even
putatively collisional blue stragglers show no connection to their
cluster environment.

What conclusions can we draw from this near-complete lack of
connection between blue stragglers and their environment? We
approached this project with the idea of looking only at blue
stragglers formed through stellar collisions (those in the core, or
the brightest blue stragglers in the core). Having found no
correlations, we are forced to acknowledge that our prediction
that core blue stragglers are predominantly formed through collisions
may be incorrect. This is in disagreement with many arguments in the
literature. Those arguments range from discussions of probable
encounter rates \citep{hills76} to the detailed dynamical simulations
of \citet{mapelli06}. It should be noted that while collisions are not
solely responsible for their production, they may still play an
important role in BSS formation. Indeed, the fraction of close
binaries has been found to be correlated with the rate of stellar
encounters in GCs \citep{pooley03}. It is becoming increasingly clear
that GCs represent complex stellar populations and that detailed
models are required in order to accurately track their evolution. 

It has also become clear, however, that blue stragglers are an elusive
bunch. It appears more and more obvious that there are numerous
factors working together to produce the populations that we
observe. Even if we limit ourselves to consider only those blue
stragglers created through binary mergers, we still need to include
the effects of cluster dynamics since the binary populations of
clusters will be modified through encounters
\citep[e.g.][]{ivanova05}. Contrary to what \cite{davies04} suggest, we
do not feel we can reliably say that the effects of collisions will be
to explicitly reduce the binary population in the core. Rather, it is
likely that the distribution of periods, separations, and mass ratios
will be modified through encounters but precisely how remains
difficult to predict. Indeed, dynamical processes act to reduce the
periods in short-period binaries rather than destroy them, with the
binaries at the lower end of the period distribution shifting to even
shorter periods the fastest \citep{andronov06}. At the same time,
wider binaries can be destroyed in clusters as a result of stellar
interactions. Therefore, models of blue straggler populations need to
be reasonably complex. In this paper, we present observational
constraints on those models. Blue straggler populations must be
approximately constant for clusters of all ages, densities,
concentrations, velocity dispersions, etc.; the number of blue
stragglers decreases with increasing cluster mass; and the type, or
luminosity function, of blue stragglers is apparently completely
random from cluster to cluster. That the luminosity function data
appears random perhaps should not have been a surprise. The current
blue straggler luminosity function is a convolution of the blue
straggler mass function and the blue straggler lifetimes.

There is one important cluster property for which we could not perform
this analysis -- the cluster binary fraction. If those clusters with
a high binary fraction also have higher relative BSS frequencies then
this might suggest a preferential tendency towards BSSs forming via
coalescence. More importantly, such a trend could be indicative of
more massive clusters having a higher frequency of binaries of the
right type. That is, more massive clusters may be more likely to
harbor binary systems with components in the right mass range and
with the right separation to form blue stragglers in the lifetime of
the cluster. It therefore seems wise to develop our knowledge of the
types of binary systems commonly found in GCs, specifically the mass
ratios, periods, and separations thereof. \cite{preston00} suggest
that GCs either destroy the primordial binaries that spawn long-period
BS binaries like those observed in the Galactic field, or they were
never home to them in the first place. This statement supports the
notion that the majority of BSSs formed in globular clusters are the
products of the mergers of close binaries, a claim that is not in
disagreement with the results of this paper. Work is required on both
the observational and theoretical fronts in order to completely
understand this ubiquitous, and frustrating, stellar population.

\acknowledgements 
We would like to thank \cite{piotto02} for providing
a robust dataset with which to explore possible BSS trends. This
research has been supported by NSERC.

\appendix

\section{Stellar Population Selection Criteria} \label{pop_selection}

Color-magnitude diagrams are often so cluttered that distinguishing
the blue stragglers from ordinary MSTO stars, or even those belonging
to the horizontal and extended horizontal branches, can be a
challenging and ambiguous task.  In looking at the BSS populations of
6 different GCs, \cite{ferraro97} defined boundaries to separate the
BSSs from regular MS stars.  Using the MSTO as a point of reference,
they shifted the CMD of each cluster to coincide with
that of M3 for consistency and then divided the blue straggler
population into two separate subsamples: bright BSSs with m$_{255}$
$<$ 19 mag and faint BSSs with 19.0 mag $<$ m$_{255}$ $<$ 19.4 mag.

Similarly, \cite{demarchi06} studied 216 open clusters (OCs)
containing a total of 2105 BSS candidates in order to compare the blue
straggler frequency to the cluster mass (total magnitude) and age.
They found an anti-correlation between BSS frequency and total magnitude,
extending the results of \cite{piotto04} to the open cluster
regime.  They also found a good correlation between the BSS frequency
and the cluster age, suggesting that at least one of the BSS formation
mechanisms requires a much longer time-scale to operate in order to
make its mark on a stellar population.  De Marchi et al. defined their
own criteria for the selection of blue stragglers and, contrary to
\cite{ferraro97}, defined boundaries by shifting the zero-age main
sequence (ZAMS) towards brighter V magnitudes and enclosed the
resident BSSs with borders above and below. 

We chose to take as our starting point the sharpest point in the bend
of the MSTO centered on the mass of points that populate it (denoted
by $((B-V)_0,V_0)$) . From here, we defined a ``MSTO width'', $w$, in
the $m_{439} - m_{555}$ plane to describe its approximate thickness,
and then established a second reference point, $(B-V, V)$, by shifting
$w/4$ mag lower in $m_{439} - m_{555}$ and $5w/8$ lower in $m_{555}$,
as indicated in Equation~\ref{eqn:shiftedMSTO}:

\begin{eqnarray}
\label{eqn:shiftedMSTO}
B-V &=& (B-V)_0 - w/4 \\
V &=& V_0 - 5w/8
\end{eqnarray}

This shift ensured that our boundary selection starting point, namely the
outer edge of the MSTO, was nearly identical for every cluster.  In order
to separate the BSSs from the rest of the stars that populate the
region just above the MSTO, we drew two lines of slope -3.5 in the
$(m_{439} - m_{555}, m_{555})$-plane made to intersect points shifted
from $(B-V,V)$.  One line was shifted 0.5 mag lower in $m_{439}
- m_{555}$ and 0.1 mag lower in $m_{555}$ from $(B-V,V)$,
whereas the other was shifted 2.0 mag lower in $m_{439} - m_{555}$ and
0.4 mag lower in $m_{555}$.  These new points of intersection are
shown in Equation~\ref{eqn:bssboundaries1} below:

\begin{eqnarray}
\label{eqn:bssboundaries1}
(B-V)_r &=& (B-V) - 0.5 \\
V_r &=& V - 0.1 \\
(B-V)_l &=& (B-V) - 2.0 \\
V_l &=& V - 0.4
\end{eqnarray}

Two further boundary conditions were defined by fitting two lines of slope
5.0, chosen to be more or less parallel to the fitted ZAMS, one
intersecting a point shifted 0.2 mag higher in $m_{439} - m_{555}$ and
0.5 mag lower in $m_{555}$ from $(B-V,V)$ (top), and another passing
through a point 0.2 mag lower in $m_{439} - m_{555}$ and 0.5 mag
higher in $m_{555}$ (bottom).  These new intersection points are given
by:

\begin{eqnarray}
\label{bssboundaries2}
(B-V)_{t, b} &=& (B-V) \pm 0.2 \\
V_{t, b} &=& V \mp 0.5
\end{eqnarray}

These cuts eliminate obvious outliers and further distinguish
BSSs from HB and EHB stars.  The methodology used in \cite{demarchi06}
for isolating BSSs similarly incorporated the ZAMS, though their upper
and lower boundaries were ultimately defined differently, as
outlined in Section~\ref{intro}.  Finally, one last cut was made to
distinguish BSSs from the EHB, namely a vertical one made 0.4 mag
lower in $m_{439} - m_{555}$ than the MSTO.

A similar methodology was used to isolate the RGB stars in the cluster
CMDs. We are restricting ourselves to RGB stars in the same magnitude
range as the blue straggler stars. First, two lines of slope -19.0 in
the $(m_{439} - m_{555}, m_{555})$-plane were drawn to intersect the
MSTO.  In order to place them on either side of the RGB, both lines
were shifted 0.6 mag lower in $m_{555}$ though it was necessary to
apply different color shifts.  The left-most boundary was placed 0.15
mag higher in $m_{439} - m_{555}$ while the right-most boundary was
placed 0.45 mag higher in $m_{439} - m_{555}$. To fully define our RGB
sample, a lower boundary was defined using a horizontal cut made 0.6
mag above the MSTO (lower in $m_{555}$). The upper boundary, on the
other hand, was simply defined to be the lower boundary of the HB.
RGB stars must therefore simultaneously satisfy:

\begin{equation}
\label{rgbboundaries}
-19.0(B-V)_{RGB} + {(V - 0.6) + 19.0((B-V) + 0.15)} < V_{RGB}
\end{equation}
\begin{equation}
V_{RGB} < -19.0(B-V)_{RGB} + {(V - 0.6) + 19.0((B-V) + 0.45)}
\end{equation}
\begin{equation}
V - 0.6 < V_{RGB}
\end{equation}
\begin{equation}
V_{RGB} > h + 0.4
\end{equation}

Pulling out HB and EHB stars from the cluster CMDs was found to be more
difficult than for BSSs and RGB stars due to the awkward shape of the
bend as the HB extends down to lower luminosities.  Exactly where the
HB ends and the EHB begins has always been a matter of some
controversy, and so the choice of where to place the boundary was
arbitrary but was at least consistent from cluster to cluster.

A vertical boundary was placed 0.5 mag lower in $m_{439} - m_{555}$ than
the MSTO to distinguish the end of the HB from the start of the EHB.
The center-most part (in absolute magnitude) of the grouping of stars
that make up the HB, denoted by $h$, was carefully chosen by eye,
and two horizontal borders were subsequently defined 0.4 mag above and
below.  One more boundary, this time a vertical cut to help
distinguish the HB from the RGB, was set 0.3, 0.4,
or 0.5 mag higher in $m_{439} - m_{555}$ than the MSTO.  The decision
of which of these three $HB_{shift}$'s to use
was decided for each cluster based on their individual CMDs.  Every
star in the CMD that fell to the left of this vertical boundary and in
between the horizontal ones was taken to be an HB star, and every star
that wasn't already counted as a BSS or a member of the HB was then
taken to be an EHB star.  Thus, HB stars, denoted by $((B-V)_{HB},
V_{HB})$, must simultaneously satisfy:

\begin{equation}
\label{hbboundaries}
(B-V)_{HB} < (B-V) + HB_{shift}
\end{equation}
\begin{equation}
h - 0.4 < V_{HB} < h + 0.4
\end{equation}

Finally, it was necessary to eliminate any very faint EHB stars to
avoid including any white dwarfs in our sample.  Hence, a final
boundary was set 3.5 mag below the lower boundary of the HB.  This
then implies that EHB stars, denoted by $((B-V)_{EHB}, V_{EHB})$, must
satisfy:

\begin{equation}
\label{ehbboundaries1}
(B-V)_{EHB} < (B-V) - 0.5
\end{equation}
\begin{equation}
h + 0.4 < V_{EHB} < h + 0.4 + 3.5 
\end{equation}

One small region of concern for the EHB remains undefined in the CMDs,
namely the portion just above the ``left'' BSS boundary with slope
-3.5 that also falls to the right of our vertical EHB/BSS dividing
line and below the lower HB boundary.  For simplicity, we treat these
stars as belonging to the EHB, though note that our frequencies would
not have been altered by much had we taken them to be BSSs, and even
less so had we taken them to be HB stars.  As such, in addition to the
above criteria, EHB stars can also simultaneously satisfy:

\begin{equation}
\label{ehbboundaries2}
(B-V)_{EHB} > (B-V) - 0.5
\end{equation}
\begin{equation}
V_{EHB} > h + 0.4
\end{equation}
\begin{equation}
V_{EHB} < -3.5(B-V)_{EHB} + {(V - 2.0) + 3.5((B-V) - 0.4)} 
\end{equation}

In Table \ref{table:data}, we give the cluster name, the number of
BSS, HB, EHB, RGB, and core stars, as well as parameters needed to
make these selections: the width of the main sequence $w$, the
position of the MSTO, and the level of the horizontal branch.
\begin{deluxetable}{lccccccccc}
\tablecaption{Population Selection Criteria and Numbers \label{table:data}}
\tablehead{
\colhead{Cluster} & \colhead{N$_{BSS}$} & \colhead{N$_{HB}$} &
\colhead{N$_{EHB}$} & \colhead{N$_{RGB}$} & \colhead{N$_{core}$} & \colhead{$w$} & \colhead{(B-V)$_{MSTO}$} & \colhead{V$_{MSTO}$} & \colhead{V$_{hb}$}
}
\startdata
NGC0104&    35&   200&     2&   648& 28924&   0.150&   0.510&  17.100&  13.800\\
NGC0362&    41&   118&    14&   343& 20359&   0.140&   0.390&  18.300&  15.200\\
NGC1261&     6&    21&     1&    51&  9265&   0.220&   0.410&  19.600&  16.700\\
NGC1851&    13&    43&     5&   117& 21923&   0.200&   0.480&  18.900&  16.000\\
NGC1904&    25&    23&    54&   238& 14485&   0.200&   0.450&  19.500&  16.000\\
NGC2808&    47&   212&   132&   975& 46328&   0.250&   0.400&  18.700&  15.500\\
NGC3201&    14&    20&     2&    83&  3175&   0.200&   0.530&  17.100&  14.100\\
NGC4147&    16&    14&     7&    53&  2675&   0.180&   0.400&  20.000&  16.900\\
NGC4372&    11&    11&     7&    95&  1847&   0.170&   0.430&  17.700&  14.500\\
NGC4590&    24&    30&     2&   180&  5253&   0.150&   0.380&  18.800&  15.500\\
NGC4833&    20&    50&    51&   300&  6461&   0.180&   0.400&  17.800&  14.500\\
NGC5024&    28&    69&    85&   421& 12997&   0.250&   0.370&  20.000&  16.700\\
NGC5634&    27&    54&    44&   252&  6868&   0.300&   0.350&  20.800&  17.500\\
NGC5694&    17&     5&   133&   184& 14914&   0.290&   0.460&  21.300&  17.800\\
NGC4499&    23&    46&     1&   186&  3221&   0.250&   0.390&  20.100&  16.900\\
NGC5824&    34&    52&   139&   233& 28046&   0.310&   0.400&  21.100&  18.000\\
NGC5904&    16&    69&    38&   270& 14696&   0.190&   0.410&  18.100&  15.000\\
NGC5927&    33&   117&     1&   324& 15856&   0.300&   0.630&  18.700&  15.200\\
NGC5946&     1&     7&    49&    89&  7032&   0.310&   0.520&  19.000&  15.500\\
NGC5986&    21&    68&   150&   588& 16141&   0.210&   0.430&  18.900&  15.600\\
NGC6093&    27&    22&   132&   356& 11390&   0.300&   0.520&  18.800&  15.400\\
NGC6171&    14&    24&     3&    93&  2972&   0.220&   0.670&  17.900&  14.600\\
NGC6205&    14&    25&   168&   735& 13276&   0.300&   0.430&  18.300&  14.700\\
NGC6229&    38&    86&    79&   385&  8999&   0.330&   0.450&  21.100&  18.000\\
NGC6218&    26&     5&    34&   166&  5142&   0.170&   0.480&  17.500&  13.900\\
NGC6235&     7&     3&    22&   106&  3288&   0.230&   0.420&  19.100&  15.500\\
NGC6266&    24&    76&    80&   483& 21369&   0.300&   0.510&  17.900&  14.700\\
NGC6273&    32&    21&   178&   715& 32692&   0.310&   0.490&  18.300&  15.000\\
NGC6284&     4&    11&    28&    74&  7890&   0.220&   0.500&  19.600&  16.400\\
NGC6287&    14&    25&    26&   132&  4827&   0.300&   0.520&  18.700&  15.400\\
NGC6293&     5&     5&    24&    52& 16707&   0.180&   0.370&  18.400&  15.200\\
NGC6304&    15&    51&     2&   162& 11524&   0.230&   0.580&  18.000&  14.600\\
NGC6342&     7&     8&     0&    34&  4809&   0.310&   0.590&  18.800&  15.400\\
NGC6356&    23&   187&     2&   565& 18223&   0.320&   0.580&  20.000&  16.500\\
NGC6362&    23&    37&     2&   161&  4084&   0.160&   0.510&  18.500&  14.900\\
NGC6388&    64&   479&    54&  1054& 46933&   0.500&   0.590&  19.100&  15.700\\
NGC6402&    19&   181&   145&   445& 12347&   0.380&   0.510&  18.700&  15.400\\
NGC6397&     4&     0&     4&     3& 16507&   0.100&   0.370&  15.700&  12.500\\
NGC6522&     0&     3&     5&    28& 16426&   0.210&   0.490&  18.600&  15.100\\
NGC6544&     3&     0&     2&     8&  3196&   0.230&   0.530&  16.300&  12.700\\
NGC6584&    22&    51&     3&   291&  5679&   0.220&   0.400&  19.500&  16.100\\
NGC6624&    15&    25&     0&   104& 12537&   0.300&   0.580&  18.600&  15.200\\
NGC6638&    18&    51&     5&   238&  6737&   0.250&   0.530&  18.900&  15.500\\
NGC6637&    20&    82&     1&   294&  8523&   0.260&   0.570&  18.800&  15.400\\
NGC6642&    14&    26&     1&    56&  2938&   0.200&   0.510&  18.600&  15.200\\
NGC6652&    14&    16&     4&    64&  5749&   0.150&   0.560&  19.000&  15.600\\
NGC6681&     4&     2&     4&    24&  6056&   0.170&   0.450&  18.800&  15.400\\
NGC6712&    39&    64&     3&   294&  9540&   0.250&   0.500&  18.100&  14.800\\
NGC6717&    10&     3&     1&    17&  1623&   0.220&   0.470&  18.500&  15.400\\
NGC6723&    14&    72&    25&   290&  7052&   0.180&   0.510&  18.600&  15.300\\
NGC6838&    17&     9&     3&    56&  2132&   0.150&   0.570&  17.000&  13.800\\
NGC6864&    27&   165&    14&   391& 10132&   0.230&   0.440&  20.200&  17.200\\
NGC6934&    18&    56&    11&   201&  9472&   0.190&   0.430&  19.700&  16.600\cr
NGC6981&    18&    50&     1&   149&  5704&   0.170&   0.390&  19.800&  16.800\cr
NGC7078&    10&    47&    32&   223& 32581&   0.210&   0.380&  18.600&  15.500\cr
NGC7089&     5&    28&    69&   244& 11723&   0.190&   0.390&  19.000&  15.500\cr
NGC7099&    10&     8&    56&    30&  8010&   0.200&   0.380&  18.300&  15.000\cr
\enddata
\end{deluxetable}


\section{Erratum:  ``Where the Blue Stragglers Roam: Searching for a Link Between Formation and Environment'' (ApJ, 661, 210 [2007]); published May 1, 2008}
%

The paper `Where the Blue Stragglers Roam: Searching for a Link
Between Formation and Environment' was published in ApJ, 661, 210
(2007).  We made an error in our conversion of arcseconds
to pixels when calculating the core radii for the WF chips.  Using a
conversion factor of 0.046 arcseconds/pixel led to an over-estimation
of the size of the core in pixels by a factor slightly exceeding 2.
Below, we provide a corrected version for each figure shown in our
original paper, in addition to corrected versions of both tables.  The
overall trends are unaffected and our figures remain similar, apart
from a scale change in Figure~\ref{fig:BSS_coll_rgb_round2} which is a
result of the error made in the calculation of our core radii.  The
new Spearman correlation coefficients also remain similar.
  
We also take this opportunity to note that the accuracy of Piotto et
al.'s (2002) assumption that each cluster core has been centered on
the PC chip proved to be more of a concern when considering these
smaller radii since for clusters with small $r_c$, the core
could be offset from the center of the PC chip by more than one core
radius.  For this erratum, we calculated our own cluster
centers by binning the stellar positions on the PC chip, fitting
Gaussians to the corresponding histograms in both the x and y
directions, and then using the peak of the distributions to extract
the cluster centers (available upon request).  The pixel coordinates
are given such that the lower left corner of the WF3 chip is at (x,y)
= (0,0).  The center of the PC chip is then at (984,984), in units of
WF chip pixels.

We present the corrected counts of blue stragglers and other
populations in Table 2, and present our revised figures in this
erratum.  The results from our previous paper remain essentially
unchanged.

\setcounter{figure}{1}

\begin{figure}
\centering
\includegraphics[width=\columnwidth]{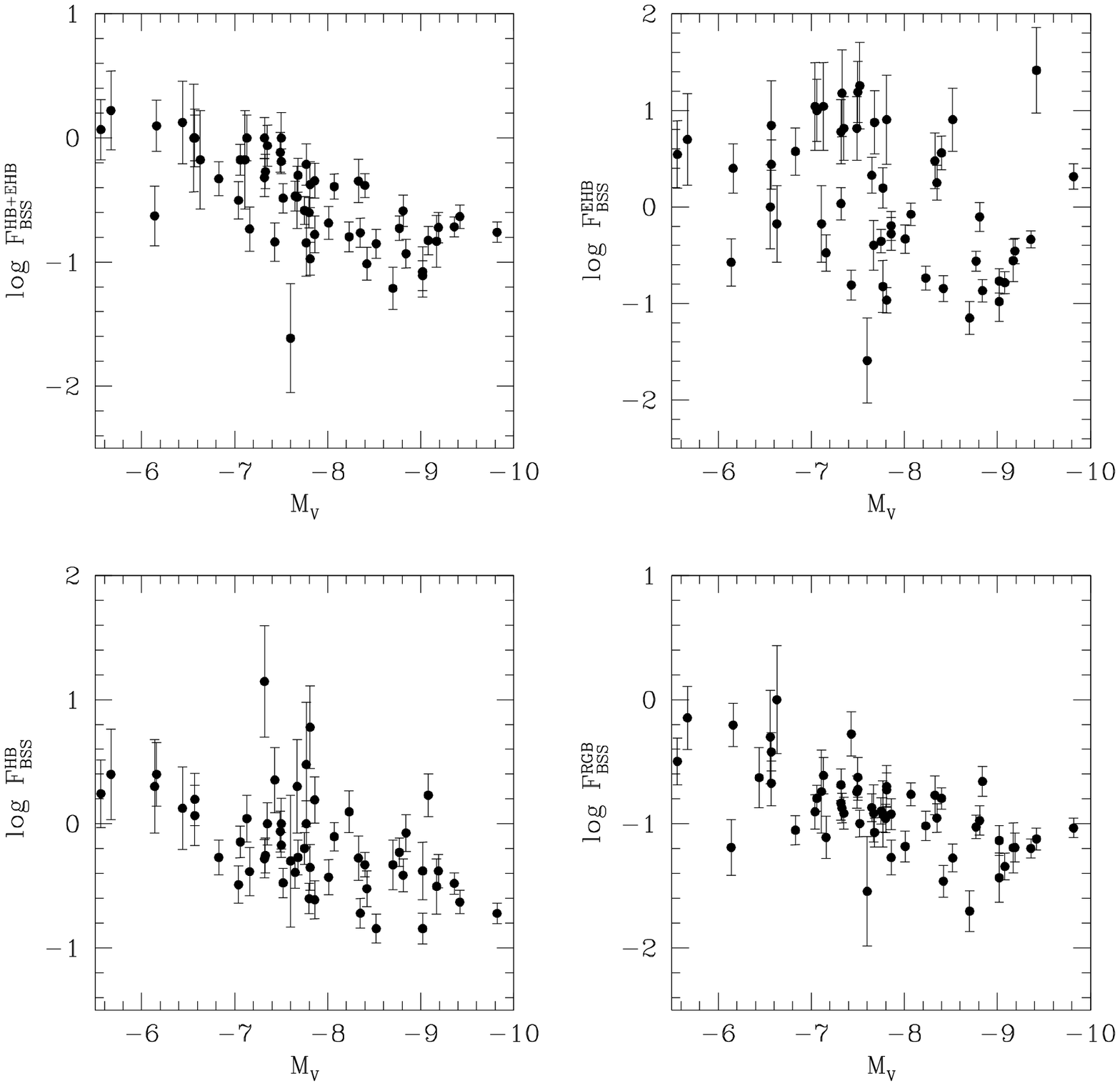}
\caption[Mv versus BSS Frequency]{Plots of the core BSS frequency
  versus the total cluster V magnitude.  Frequencies were normalized
  using RGB stars (bottom right), HB stars (bottom left), EHB stars
  (top right), and finally, HB \& EHB stars combined (top left). 
\label{fig:BSSnorm_round2}}
\end{figure}

\begin{figure}
\centering
\includegraphics[width=\columnwidth]{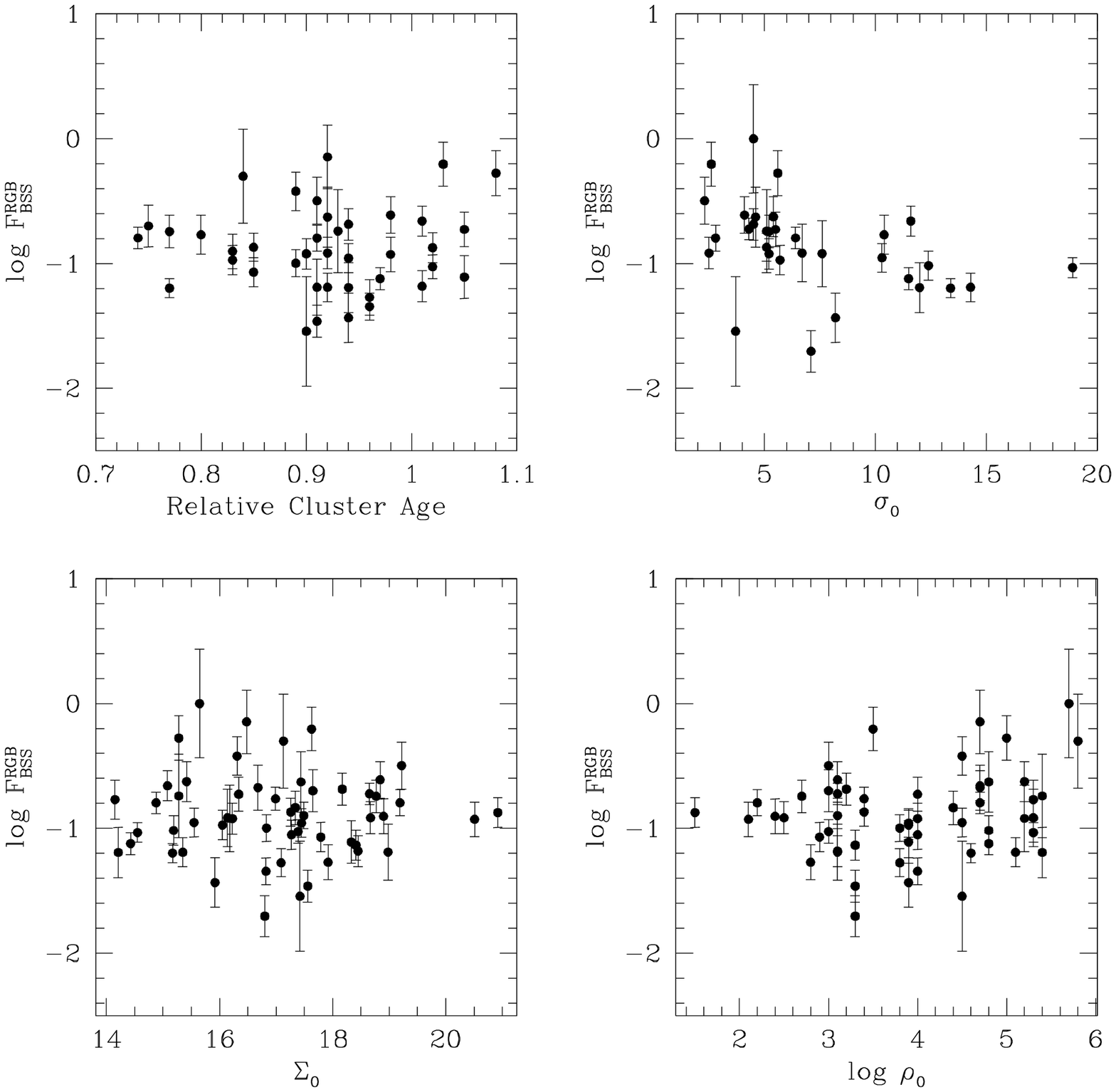}
\caption[Parameters versus BSS Frequency]{Plots of the core BSS frequency
  versus the logarithm of the central density (bottom right), the
  central surface brightness (bottom left), the relative cluster age
  (top left), and the central velocity dispersion (top right).
  Frequencies were normalized using RGB stars.  The central density is
  given in units of L$_{\odot}$ pc$^{-3}$, the central surface
  brightness in units of V mag arcsecond$^{-2}$, and the central
  velocity dispersion in units of km s$^{-1}$.  The cluster age is
  normalized, however, and its values represent the ratio between
  the cluster age and the mean age of a group of metal-poor clusters
  as described in \cite{deangeli05}.  
\label{fig:BSSresults_round2_rgb}}
\end{figure}

\begin{figure}
\centering
\includegraphics[width=\columnwidth]{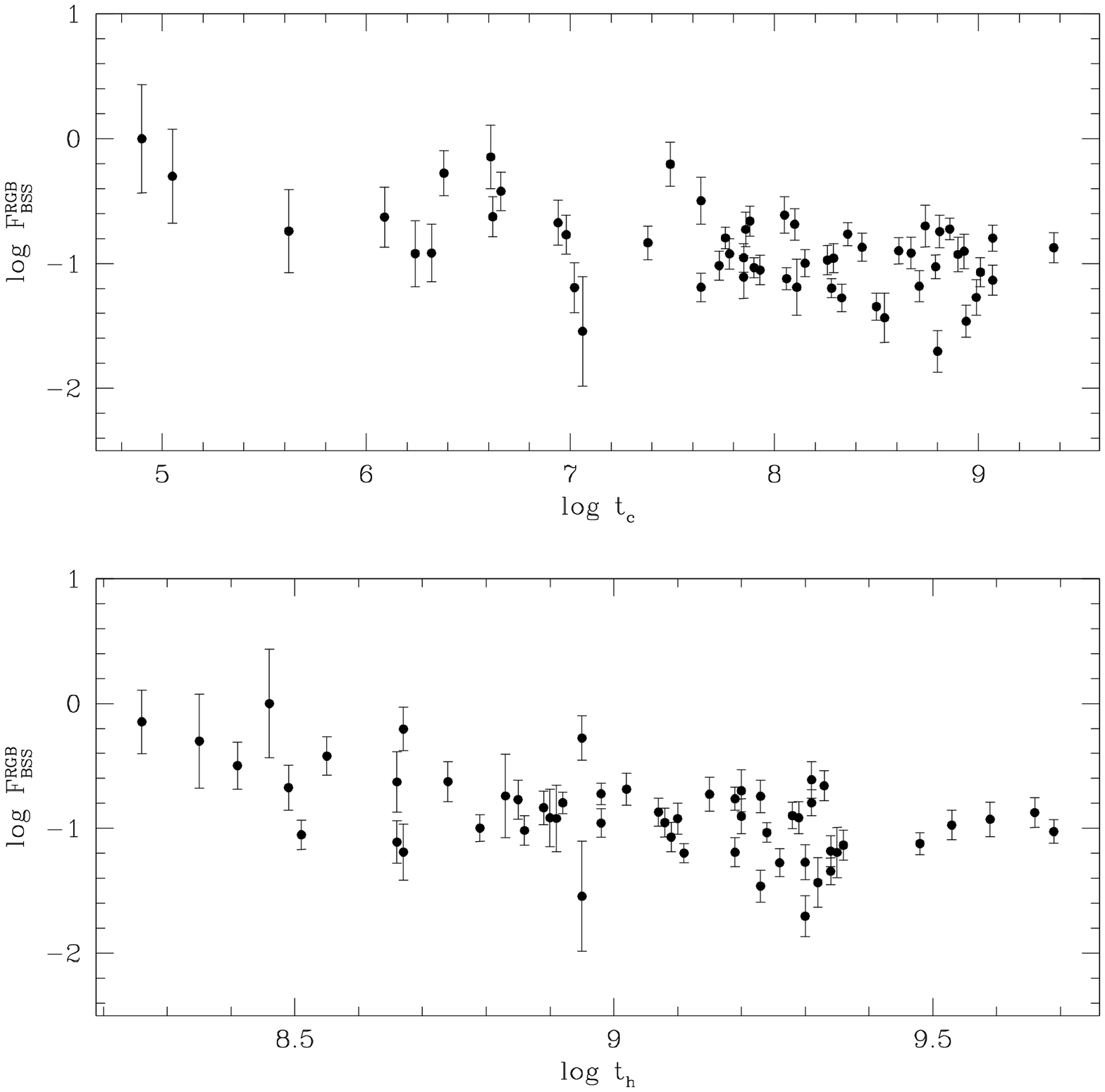}
\caption[Log of Core Relaxation Time and at the Half Mass-Radius versus BSS Frequency]{Plots of the core BSS frequency
  versus the logarithm of the core relaxation time in years (top), and
  the logarithm of the relaxation time at the half-mass radius in
  years (bottom).  Frequencies were normalized using RGB stars.  Note
  the anti-correlation that exists between F$_{BSS}$ and log $t_h$.
\label{fig:BSS_th_tc_rgb_round2}}
\end{figure}

\begin{figure}
\centering
\includegraphics[width=\columnwidth]{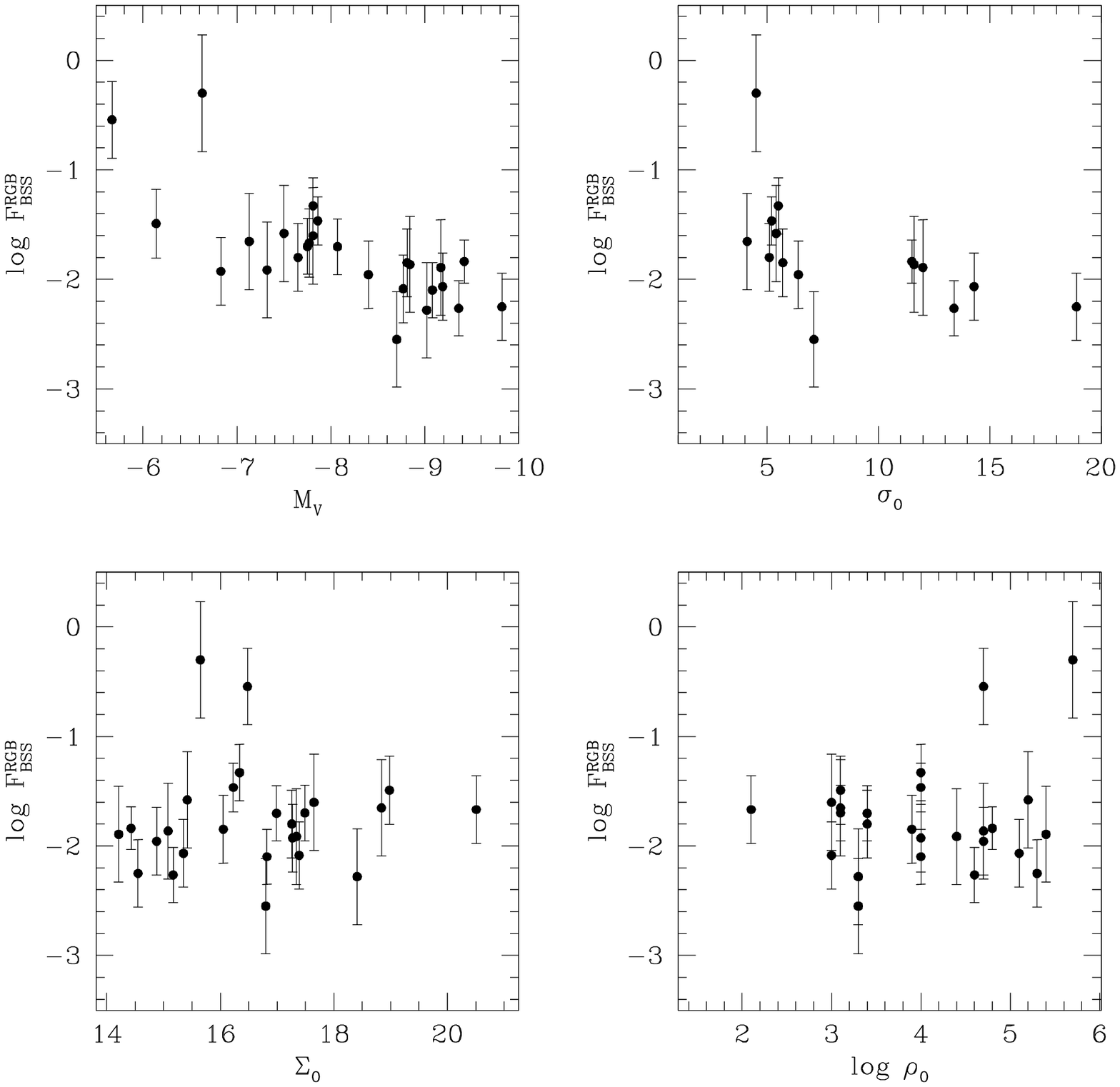}
\caption[Parameters Versus Brightest BSSs]{Plots of the brightest core
  BSS frequency versus the logarithm of the central density (bottom
  right), the central surface brightness (bottom left), the total
  cluster V magnitude (top left), and the central velocity dispersion
  (top right).  Frequencies were normalized using RGB stars.  The
  central density is given in units of L$_{\odot}$ pc$^{-3}$, the
  central surface brightness in units of V mag arcsecond$^{-2}$, the
  cluster magnitude in V mag, and the central velocity dispersion in
  units of km s$^{-1}$.
\label{fig:BBSS4x4_round2}}
\end{figure}

\begin{figure}
\centering
\includegraphics[width=\columnwidth]{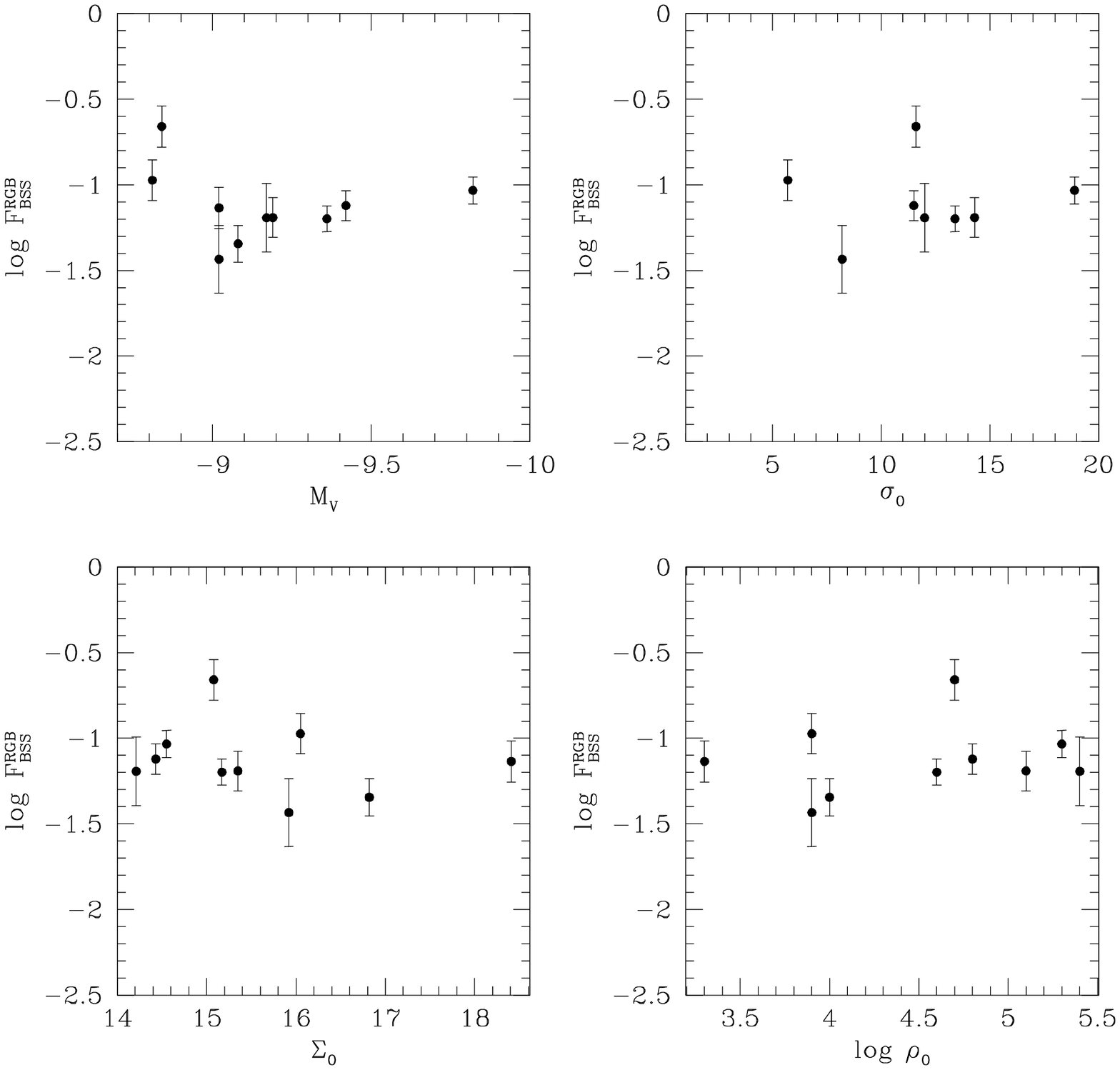}
\caption[Parameters Versus Brightest BSSs]{Plots of the core BSS
  frequency in only the brightest clusters ($M_V < -8.8$)
  versus the logarithm of the central density (bottom right), the
  central surface brightness (bottom left), the total cluster V
  magnitude (top left), and the central velocity dispersion (top
  right).  Frequencies were normalized using RGB stars.
  The central density is given in units of L$_{\odot}$ pc$^{-3}$, the
  central surface brightness in units of V mag arcsecond$^{-2}$, the
  cluster magnitude in V mag, and the central velocity dispersion in
  units of km s$^{-1}$. 
\label{fig:BSS4x4_brightestMv_round2}}
\end{figure}

\begin{figure}
\centering
\includegraphics[width=\columnwidth]{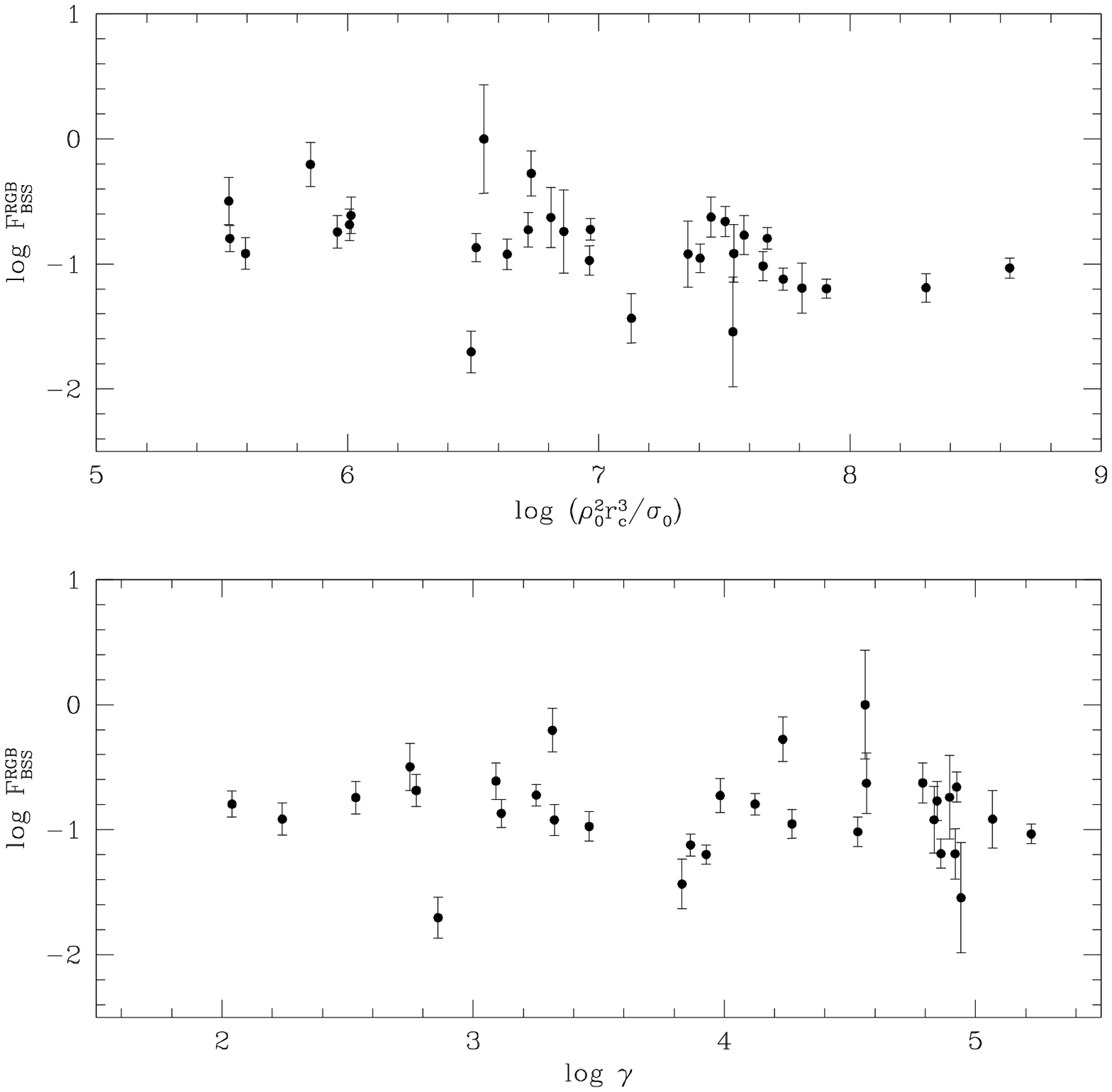}
\caption[Collisional Parameter versus BSS Frequency]{Plot of the
  BSS frequency within the cluster core versus the rate of stellar
  collisions per year using ${\Gamma} = \frac{{\rho}_0^{2}r_c^{3}}{{\sigma}_0}$ as the
  collisional parameter (top).  $\rho_0$ is the central density in units
  of L$_{\odot}$ pc$^{-3}$, $\sigma_0$ is the central velocity
  dispersion in km s$^{-1}$, and $r_c$ is the core radius in
  parsecs.  BSS frequency is also plotted against the probability of a
  stellar collision occurring within the core in one year (bottom).
  Frequencies were normalized using core RGB stars.
\label{fig:BSS_coll_rgb_round2}}
\end{figure}

\begin{deluxetable}{lcc}
\tablewidth{0pt}
\tablecaption{Spearman Correlation Coefficients \label{table:spearman}}
\tablehead{
\colhead{Parameter} & \colhead{$r_s$} & \colhead{Probability}\\
}
\startdata
Total cluster V magnitude  	  & 0.60 & 	 1.95E-06\\
Central velocity dispersion  	  &-0.56 &	 1.3E-03\\
Half-mass relaxation time         &-0.41 &	 2.1E-03\\
Core relaxation time              &-0.37 & 	 6.1E-03\\
Collision rate                    &-0.52 &	 3.2E-03\\
Surface brightness   	          & 0.09 &	 0.54\\
Central density                   & 0.03 &	 0.85\\
Collision probability             &-0.25 &	 0.18\\
Age  	                          & 0.04 &       0.80\\
\enddata
\end{deluxetable}

\begin{figure}
\centering
\includegraphics[width=\columnwidth]{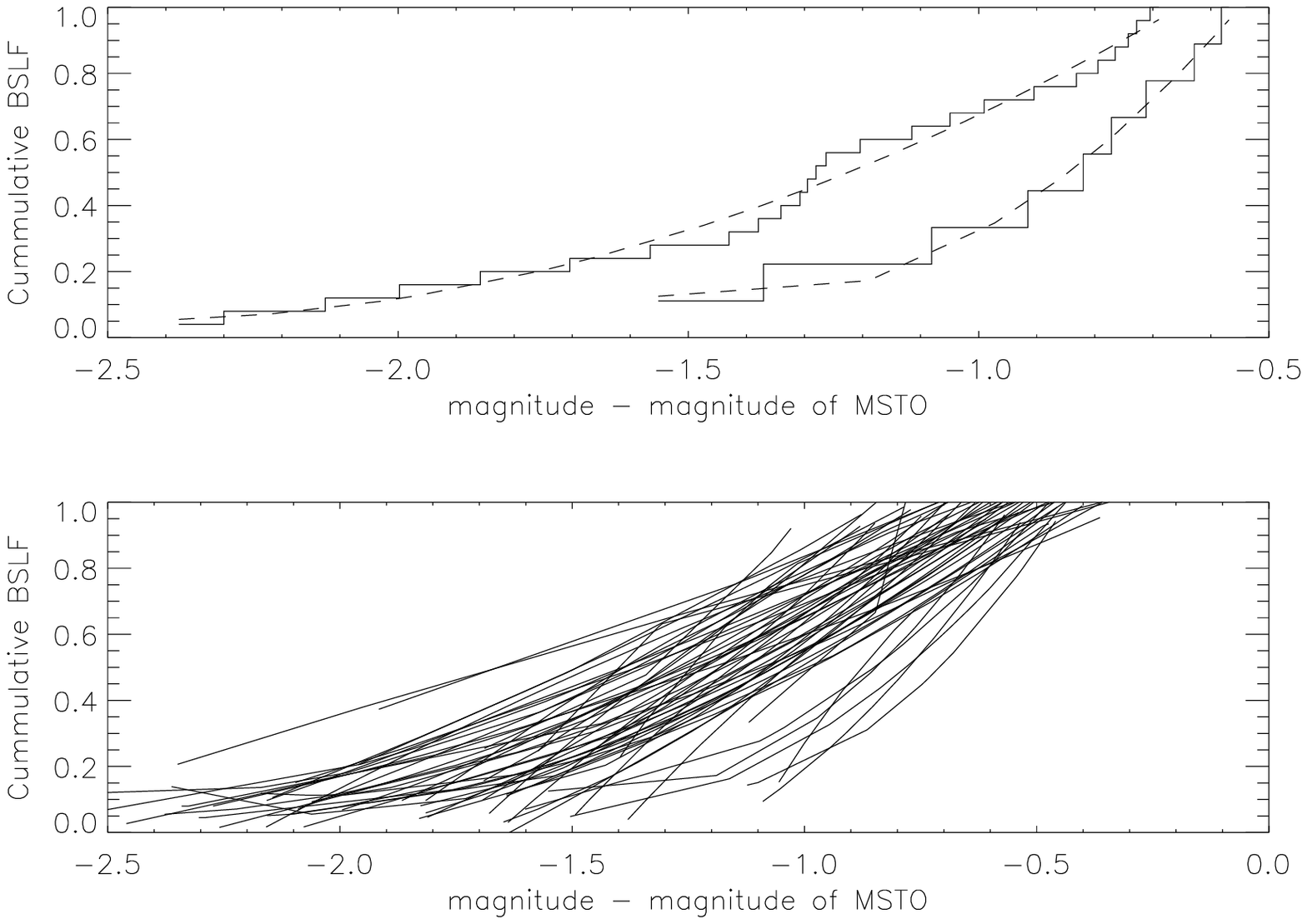}
\caption[BSLFs and fits]{Cumulative BS luminosity functions. The top
  panel shows the BSLFs for NGC 104 (47 Tucanae) and NGC 7099 (M30)
  (solid lines) along with quadratic fits to those functions (dashed
  lines). The bottom panel shows the quadratic fits to all cluster
  BSLFs. 
  \label{fig:BSS_LF}}
\end{figure}

\begin{figure}
\centering
\includegraphics[width=\columnwidth]{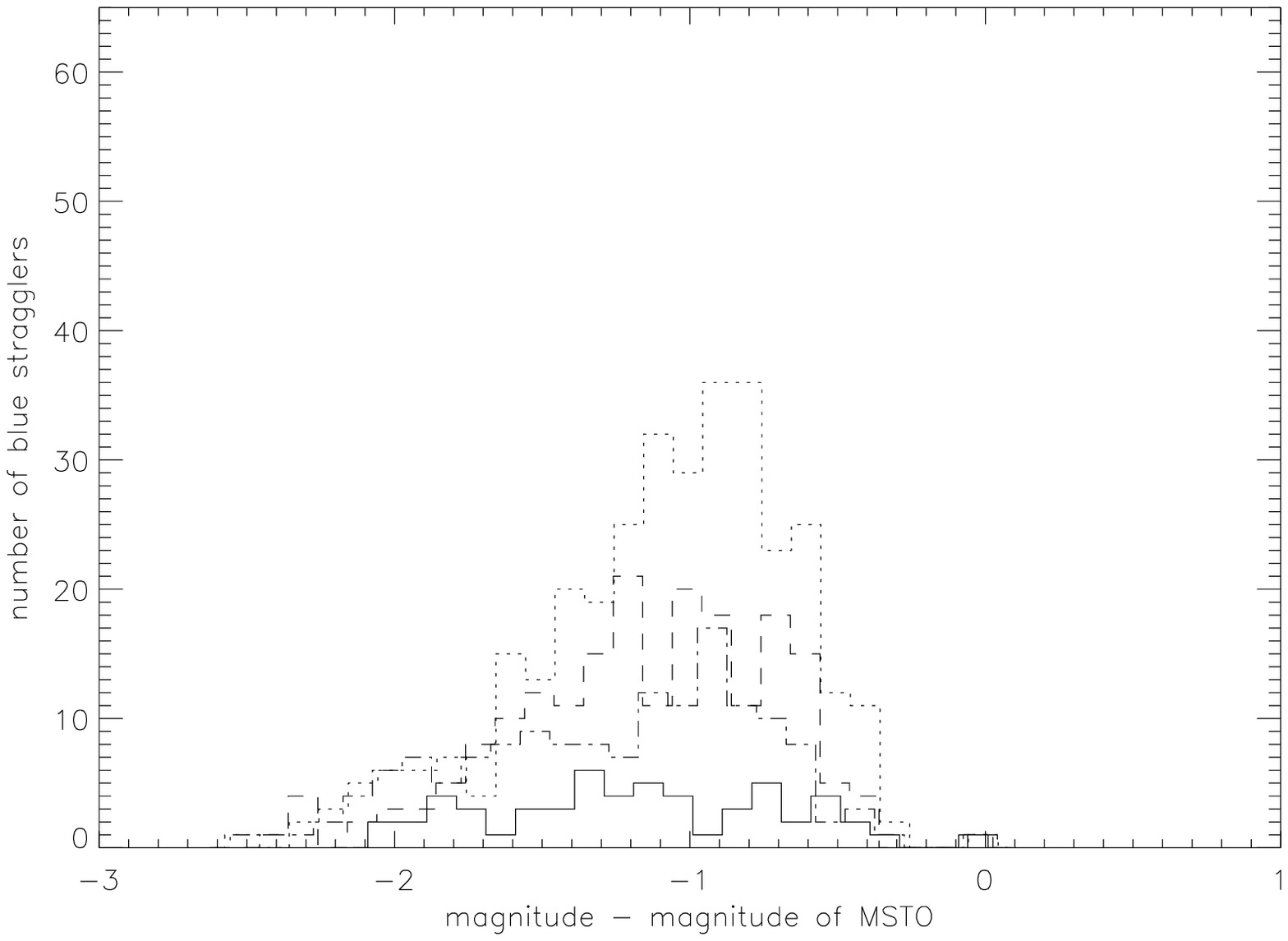}
\caption[binned BSLF with MV]{Blue straggler luminosity functions,
  binned according to total cluster magnitude. The solid line is for clusters
  with $M_V$ between -6 and -7; dotted line for clusters with $M_V$
  between -7 and -8; dashed line for clusters with $M_V$ between -8
  and -9, and the dash-dotted line is for clusters with $M_V$ between
  -9 and -10. 
\label{fig:BSLF_MV_bin}}
\end{figure}

\begin{deluxetable}{lccccccccc}
\tablewidth{0pt}
\tablecaption{Population Selection Criteria and Numbers
\label{table:data}}
\tablehead{
\colhead{Cluster} & \colhead{N$_{BSS}$} & \colhead{N$_{HB}$} &
\colhead{N$_{EHB}$} & \colhead{N$_{RGB}$} & \colhead{N$_{core}$} &
\colhead{$w$} & \colhead{(B-V)$_{MSTO}$} & \colhead{V$_{MSTO}$} & \colhead{V$_{hb}$}
}
\startdata
NGC0104&   26&   111&     1&   344&   7398&   0.150&   0.510&   17.100&   13.800\\
NGC0362&   29&    62&     8&   181&   3541&   0.140&   0.390&   18.300&   15.200\\
NGC1261&    8&    18&     1&    40&   3241&   0.220&   0.410&   19.600&   16.700\\
NGC1851&    9&    17&     3&    53&    539&   0.200&   0.480&   18.900&   16.000\\
NGC1904&   14&     9&    22&   117&   2050&   0.200&   0.450&   19.500&   16.000\\
NGC2808&   35&   106&    76&   552&   9525&   0.250&   0.400&   18.700&   15.500\\
NGC3201&   13&    15&     2&    72&   2671&   0.200&   0.530&   17.100&   14.100\\
NGC4147&   10&     4&     4&    16&    344&   0.180&   0.400&   20.000&   16.900\\
NGC4372&   11&    11&     7&    93&   1794&   0.170&   0.430&   17.700&   14.500\\
NGC4590&   13&    13&     2&   107&   2259&   0.150&   0.380&   18.800&   15.500\\
NGC4833&   13&    35&    28&   198&   3891&   0.180&   0.400&   17.800&   14.500\\
NGC5024&   23&    39&    84&   244&   3159&   0.250&   0.370&   20.000&   16.700\\
NGC5634&   19&    30&    43&   150&   1588&   0.300&   0.350&   20.800&   17.500\\
NGC5694&   12&     2&   111&    64&    544&   0.290&   0.460&   21.300&   17.800\\
IC 4499&   15&    27&     1&   112&   1755&   0.250&   0.390&   20.100&   16.900\\
NGC5824&   16&    19&   118&    73&    379&   0.310&   0.400&   21.100&   18.000\\
NGC5904&   15&    39&    19&   141&   3170&   0.190&   0.410&   18.100&   15.000\\
NGC5927&   16&    64&     0&   145&   3168&   0.300&   0.630&   18.700&   15.200\\
NGC5946&    1&     2&    39&    35&    391&   0.310&   0.520&   19.000&   15.500\\
NGC5986&   12&    40&    84&   348&   6960&   0.210&   0.430&   18.900&   15.600\\
NGC6093&   15&    12&    82&   156&   1326&   0.300&   0.520&   18.800&   15.400\\
NGC6171&   11&    10&     1&    45&    837&   0.220&   0.670&   17.900&   14.600\\
NGC6205&    7&    15&    99&   354&   4285&   0.300&   0.430&   18.300&   14.700\\
NGC6229&   26&    33&    31&   151&   1484&   0.330&   0.450&   21.100&   18.000\\
NGC6218&   14&     1&    13&    68&   1715&   0.170&   0.480&   17.500&   13.900\\
NGC6235&    4&     2&    15&    62&    928&   0.230&   0.420&   19.100&   15.500\\
NGC6266&   15&    36&    43&   233&   2760&   0.300&   0.510&   17.900&   14.700\\
NGC6273&   17&    10&   104&   376&   8015&   0.310&   0.490&   18.300&   15.000\\
NGC6284&    0&     4&     9&    21&    357&   0.220&   0.500&   19.600&   16.400\\
NGC6287&    7&    17&    21&    90&    916&   0.300&   0.520&   18.700&   15.400\\
NGC6293&    3&     1&    20&    25&    331&   0.180&   0.370&   18.400&   15.200\\
NGC6304&   12&    23&     2&    82&   1635&   0.230&   0.580&   18.000&   14.600\\
NGC6342&    4&     3&     0&    17&    175&   0.310&   0.590&   18.800&   15.400\\
NGC6356&   16&   112&     2&   302&   3817&   0.320&   0.580&   20.000&   16.500\\
NGC6362&   20&    28&     2&   125&   3104&   0.160&   0.510&   18.500&   14.900\\
NGC6388&   33&   174&    16&   356&   2593&   0.500&   0.590&   19.100&   15.700\\
NGC6402&   14&    98&    82&   191&   6513&   0.380&   0.510&   18.700&   15.400\\
NGC6397&    2&     0&     3&     2&     96&   0.100&   0.370&   15.700&   12.500\\
NGC6522&    4&     2&    10&    33&    295&   0.210&   0.490&   18.600&   15.100\\
NGC6544&    2&     0&     2&     4&     34&   0.230&   0.530&   16.300&   12.700\\
NGC6584&   15&    28&     2&   176&   2385&   0.220&   0.400&   19.500&   16.100\\
NGC6624&    9&     9&     0&    38&    454&   0.300&   0.580&   18.600&   15.200\\
NGC6638&   15&    28&     4&   169&   1915&   0.250&   0.530&   18.900&   15.500\\
NGC6637&   18&    54&     1&   179&   2681&   0.260&   0.570&   18.800&   15.400\\
NGC6642&    7&     6&     1&    33&    301&   0.200&   0.510&   18.600&   15.200\\
NGC6652&   11&     7&     4&    29&    466&   0.150&   0.560&   19.000&   15.600\\
NGC6681&    2&     0&     3&    11&     92&   0.170&   0.450&   18.800&   15.400\\
NGC6712&   31&    46&     2&   164&   5197&   0.250&   0.500&   18.100&   14.800\\
NGC6717&    5&     2&     1&     7&     97&   0.220&   0.470&   18.500&   15.400\\
NGC6723&   10&    41&    19&   187&   4246&   0.180&   0.510&   18.600&   15.300\\
NGC6838&    7&     4&     2&    22&    601&   0.150&   0.570&   17.000&   13.800\\
NGC6864&   16&    84&     9&   144&   1364&   0.230&   0.440&   20.200&   17.200\\
NGC6934&   17&    42&     8&   126&   2506&   0.190&   0.430&   19.700&   16.600\\
NGC6981&   11&    34&     1&    88&   2085&   0.170&   0.390&   19.800&   16.800\\
NGC7078&    5&    16&    18&    78&    777&   0.210&   0.380&   18.600&   15.500\\
NGC7089&    5&    12&    48&   136&   1989&   0.190&   0.390&   19.000&   15.500\\
NGC7099&    9&     4&    58&    17&    315&   0.200&   0.380&   18.300&   15.000\\
\enddata
\end{deluxetable}

\end{document}